%% file: 0MAINArxiv.tex
\RequirePackage{fix-cm}
\documentclass[twocolumn]{svjour3}          % twocolumn
\smartqed  % flush right qed marks, e.g. at end of proof
\usepackage{graphicx}

\usepackage[colorinlistoftodos]{todonotes} 
\usepackage{xargs}
\usepackage{graphicx,psfrag,epsf}
\usepackage{enumerate}
\usepackage{verbatim}
\usepackage{tabularx}
\PassOptionsToPackage{hyphens}{url}\usepackage{hyperref}
\begin{document}

\title{User Characterization for Online Social Networks%\thanks{Grants or other notes
}
\subtitle{
%Do you have a subtitle?\\ If so, write it here
}


\author{Tayfun Tuna   \and Esra Akbas   \\
 Ahmet Aksoy   \and M. Abdullah Canbaz  \\
Umit Karabiyik \and Bilal Gonen \\ Ramazan Aygun
        }
\authorrunning{T.Tuna, E.Akbas, A.Aksoy, M.A.Canbaz, U Karabiyik, B.Gonen and R.Aygun} % if too long for running head

\institute{%F. Author \at
           %   first address \\
           %   Tel.: +123-45-678910\\
           %   Fax: +123-45-678910\\
           %   \email{fauthor@example.com}           %  \\
\\
Tayfun Tuna  \at
Department of Computer Science\\
University of Houston \\
\email{ttuna@uh.edu} 
\and
Esra Akbas\at
Department of Computer Science \\ 
Florida State University \\
\email{akbas@cs.fsu.edu} 
 \and  
Ahmet Aksoy  and M. Abdullah Canbaz \at
Department of Computer Science and Engineering \\
University of Nevada, Reno \\
\email{aksoy@nevada.unr.edu, canbaz@unr.edu} 
 \and  
Umit Karabiyik \at
Department of Computer Science \\
Sam Houston State University\\
\email{umit@shsu.edu} 
 \and  
Bilal Gonen \at
School of Information Technology  \\ 
University of Cincinnati\\
%University of West Florida \\
\email{bilal.gonen@uc.edu} 
%850-483-0444 \\
 \and  
Ramazan Aygun \at
Department of Computer Science \\
University of Alabama in Huntsville \\
\email{aygunr@uah.edu} 
%2568246455 \\
}
\date{Copyright © 2016, Springer-Verlag Wien, Social Network Analysis and Mining.}


\maketitle

\begin{abstract}
Online social network analysis has attracted great attention with a vast number of users sharing information and availability of APIs that help to crawl online social network data. In this paper, we study the research studies that are helpful for user characterization as online users may not always reveal their true identity or attributes. We especially focused on user attribute determination such as gender, age, etc.; user behavior analysis such as motives for deception; mental models that are indicators of user behavior; user categorization such as bots vs. humans; and entity matching on different social networks. We believe our summary of analysis of user characterization will provide important insights to researchers and better services to online users.

\end{abstract}

\input{1introduction.tex}

\input{2identifying_user_attributes.tex}

\input{3user_behavior.tex}
\input{4mental_models.tex}

\input{5user_categorization.tex}

\input{6matching_profiles.tex}

\input{7conclusion.tex}

\bibliographystyle{spmpsci}
\bibliography{references}
%\begin{thebibliography}{}
% etc
%\end{thebibliography}

\end{document}

%% file: 1introduction.tex
\section{Introduction}
\begin{figure}[!htb]
\includegraphics[width=0.5\textwidth,keepaspectratio]{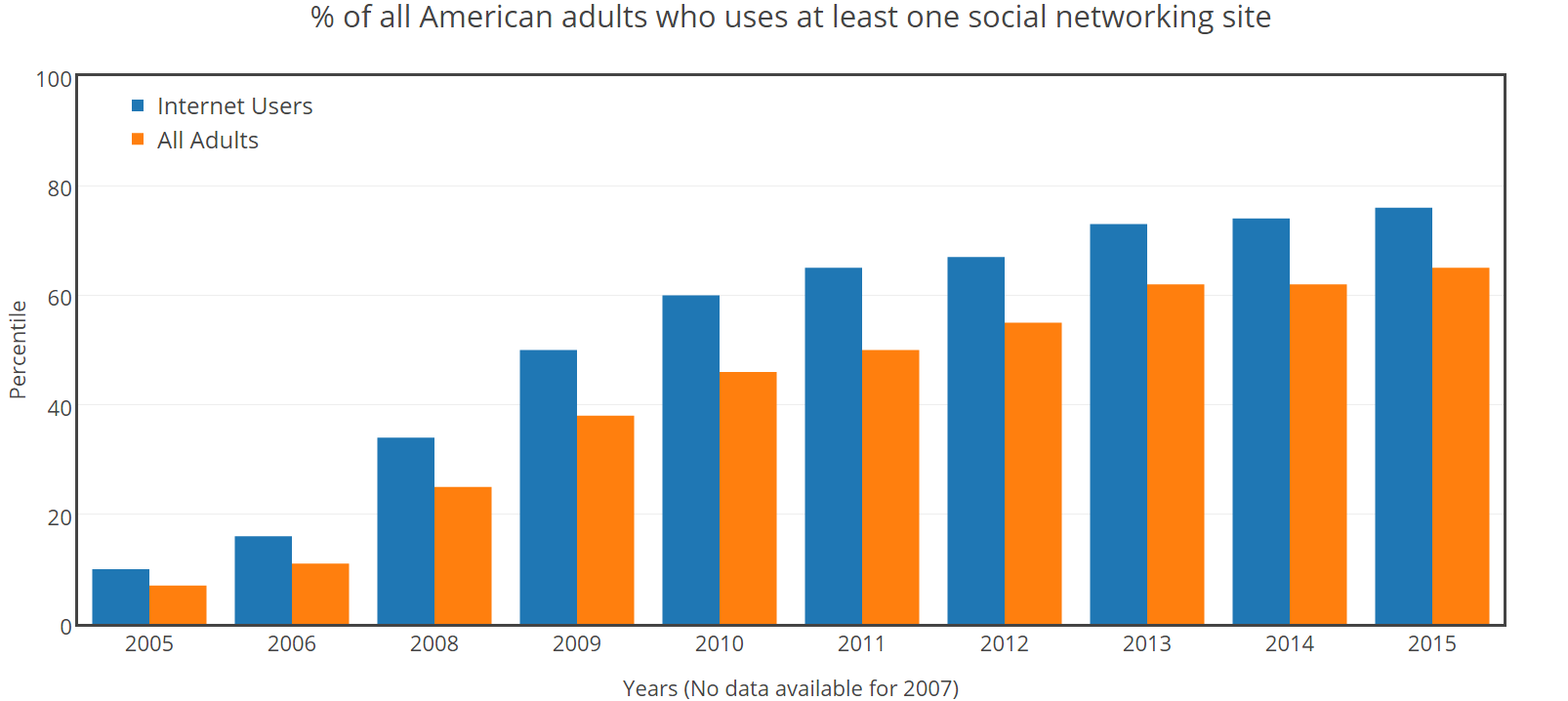}
\caption{Social networking site usage for all adults and internet-using adults}
\label{fig:SocialMediaUsage}
\end{figure}

\begin{figure}[!htb]
\centering
\includegraphics[width=0.5\textwidth,keepaspectratio]{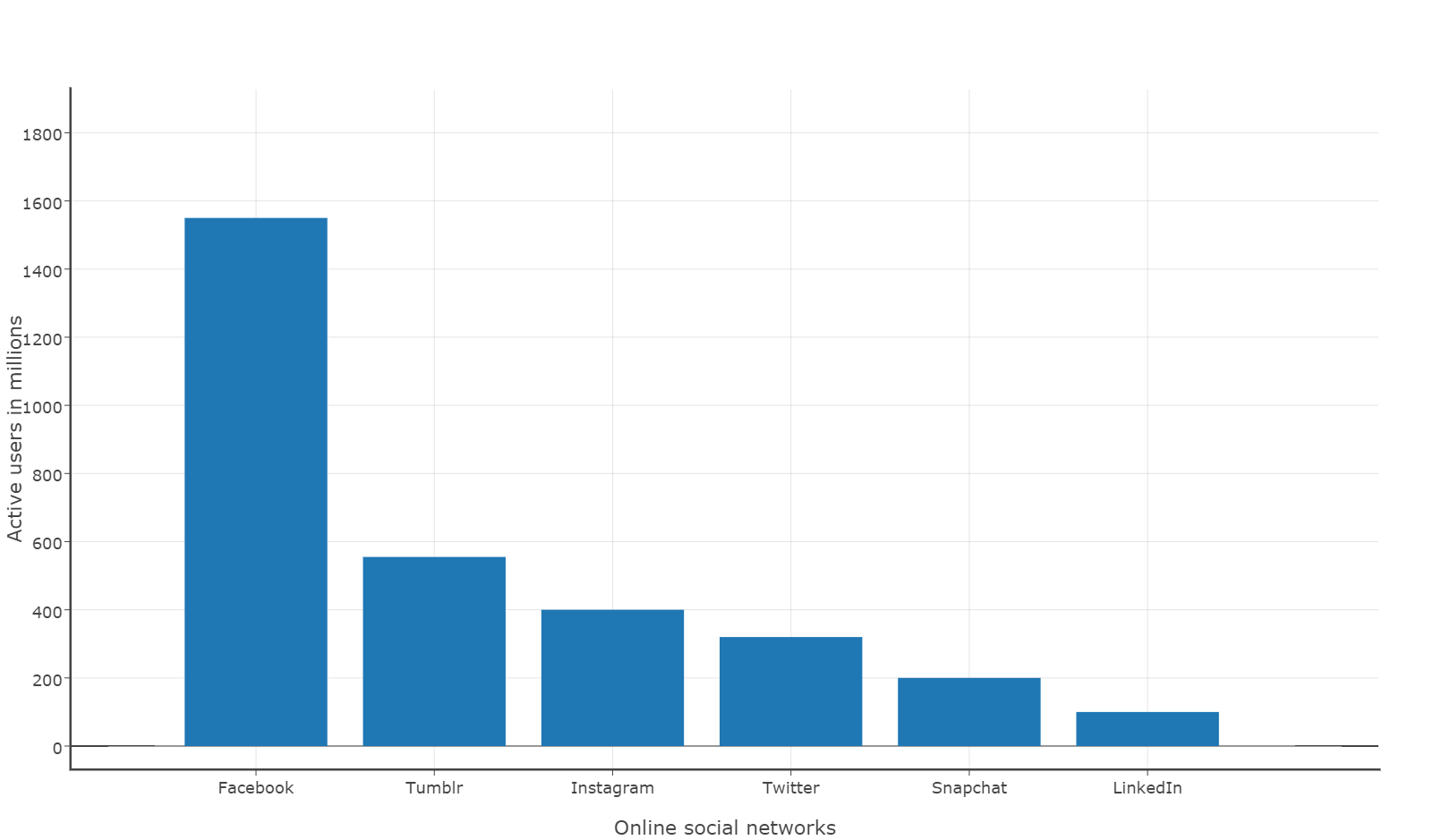}
\caption{Active users of social networks worldwide as of January 2016}
\label{fig:statsOSNs}
\end{figure}

Online social networking is one of the recent developments that attract everyone regardless of their age, gender, socioeconomic status, etc. and produces tremendous amount of digital data for analysis \cite{Raghavan-13}. The number of online social network users is increasing every year  as the hand-held mobile devices  become part of our every day life. According to the Pew Research Center's survey report in \cite{SocialMediaUsage}, 65\% of adult Americans use at least one social networking site such as Facebook, Instagram, Twitter, LinkedIn, and so on. Figure \ref{fig:SocialMediaUsage} shows the change in social networking site usage in the last decade where Figure \ref{fig:statsOSNs} shows the number of users worldwide in millions for some of the well known Online Social Networks (OSNs) \cite{OSNusers}.

Social network analysis (SNA) is a method of using several methods in network and graph theories in order to study and analyze social structures for different circumstances \cite{Otte-02}. The vast available data can be studied and interpreted from different perspectives. For example, from the law enforcement point of view, the main goal of using SNA would be to understand the relations (links or edges) between the actors (nodes) for a variety of purposes such as solving criminal activities, preventing terrorist attacks, detecting deceptions, etc. Since perpetrators' real personas are likely to be unknown and hidden behind the screens in many cyber-crimes, understanding the characteristics and behavioral dynamics of criminals has been critical for law enforcement officers  to identify the real persons. Even people with many online personas (e.g., Joshua Goldberg \cite{Zavadski-15}) can be easily characterized using a variety of techniques including identifying user attributes, analyzing user behavior, studying mental models, etc. In this paper, we study the research efforts which mainly focus on user characteristics for online social networks. These research efforts play critical role to understand several dynamics and activities of OSN account holders.

Online social networks are probably one of the major platforms for supporting freedom of thoughts as well as free speech. One of the major reasons for this is that the privacy of users could be maintained at a level or users do not need to reveal true identities such as their real name, age, gender, profession, and location. This also brings some challenges for typical users of online social networks or marketing people. To market the correct product to the correct people or to know actual attributes of the person being communicated has been more difficult since the people behind their systems are not visible. However, the profiles of users, the languages used by users, frequency of messages, the number of friends, the number of common friends, etc. can be determined  from user profiles, networks of users, and contents of messages. Such information could be used to identify the characteristics of users. In this paper, we provide an overview of attribute determination, behavior analysis, mental models, user categorization, and entity resolution.

There have been some research studies including surveys aiming to identify, understand, characterize and measure user behaviors in OSNs \cite{maia-08} \cite{jin-13} \cite{Benevenuto-09} \cite{gyarmati-10}. These research studies are conducted primarily for the purpose of modeling the user behaviors for marketing, advertising, subscription and membership perspectives. However, our work aims to present the studies from different perspectives such as finding hidden information from posted user data, determining features for characterization and user behavior analysis as well as identifying user trustworthiness, mistrust, speciousness and maliciousness. We also believe that our study on user characterization will provide better services for online users and secure environment for information exchange.

We also present and summarize the researchers' findings on their social network analysis which will help us understand the dynamics of social networks and to answer questions as follows. “Do online social network users resist sharing their information for security reasons? Can we measure a person's behavior: openness, curiosity, conscientiousness, neuroticism? How do we measure the privacy of a user and how can online social networks guarantee users' happiness and trust while sharing data? Does it become a habit for a person to share content in social network or is it possible to predict whether a person will tweet in a specific time? Do spammers affect trending topics? What reactions do social network users have for social attacks?” Our goal in this paper is to present these diverse issues of user characterization in an organized and concise manner.

This paper is organized as follows. The following section provides methods of user attribute determination such as gender, age, geo-location, and prediction. Section \ref{sec:userbehavior} provides different types of user behaviors and impact of social networks on user behaviors such as privacy, deception, radicalism, etc. The mental models of online social user models are discussed on Section \ref{sec:mentalmodels}. Section \ref{sec:usercategorization} provides methods of distinguishing real users from spammers, fake users and bots. Matching profiles of the same user on different online social networks is discussed in Section \ref{sec:entitymatching}. Finally, the last section summarizes the studies with overview and lists the open research issues before the conclusion.

%% file: 2identifying_user_attributes.tex
\section{Identifying User Attributes}

User attribute identification is an important task for many different applications such as personalized recommendation and targeted advertising. Thus, a number of studies have been performed to identify user attributes automatically from SNS (social network services). While some of them use postings that include various attributes of writers, some of them use only profile information of writers. There are also some studies that use both of them. In these studies, the goal of user characterization is detecting different features of users such as gender, age, occupation, and interest.

\subsection{Gender Detection}

It is important to know the gender of the users in many fields. For example, recommendation or advertising can be done based on the gender. Also, in social networks, gender detection can be used to detect fake user accounts. While some studies use text messages of users to detect gender, some studies user profile information for this purpose.

While writing a text, there are differences in  ways that men and women use the language. Deitrick et al. \cite{Deitrick-2012} firstly extract 9170 features (95 1-gram characters available on a keyboard, 9025 2-gram features, and 50 features from \cite{Burger-2011}) from 3,031 tweets (one tweet per user) after filtering 36,238 tweets. They have reduced the number of features to 53 using WEKA's feature selection tools: relief, information gain, information gain ratio, symmetrical uncertainty, chi-square, and filtered attribute evaluation. Using modified balanced Winnow neural network, they are able to predict gender with an accuracy of 98.51\% and precision of 97.98\%. 

Besides using n-grams, linguistic features can also be used to predict the gender of an author. Cheng et al. \cite{cheng-11} aim to identify the gender of the author of a given text document which could be short, multi-genre and context free by analyzing the linguistic features of text entrees. A feature space composed of 5 sets of gender related features is constructed: 1) character-based, (2) word-based, (3) syntactic, (4) structure based, and (5) function words. From two sources, total 545 features are selected and stored in the feature space. The first source is from Reuter’s newsgroup dataset \cite{lewis-04}, which includes English stories written by journalists.  The length of stories varies from 200 to 1000 words. The second source is \textit{Enron email data set} \cite{klimt-04}, which includes emails of employees. Each text is represented by selected features. 
Models are trained using Bayesian-based Logistic Regression, Decision Tree, and Support Vector Machine (SVM). The best classification result is obtained by SVM with the accuracy of 76.75\% (Reuters newsgroup) and 82.23\% (Enron Corpus) on these datasets. This result implies the fact that classifying genders of the authors of neutral news (Reuters Newsgroup) is more difficult than classifying genders of the authors of personal e-mails (Enron Corpus). Another drawn conclusion is that function words, word-based features and structural features are strong gender discriminators. 

Using text for gender analysis may generate a huge number of features, which makes gender classification computationally expensive. Alowibdi et al. \cite{alowibdi-13-2} propose a method to reduce the feature set from several million features for gender classification on Twitter. Gender is identified based on only user profiles using three characteristics: (1) first name, (2) user name, and (3) profile colors. Profile colors consist of the background, text, link, sidebar fill, and sidebar border colors. A large dataset is created from Twitter profiles to evaluate their methods. The ground truth of a user's gender is found by following the links from the profiles to other Online Social Networks (OSN) having gender information. The characters of different languages are converted to characters in English language using the Google Input Tools (GIT). After converting  first names and user names to phoneme sequences, classifiers are trained using their feature set by applying Naive Bayes (NB), Decision Tree (DT) and Naive-Bayes-Decision-tree (NB-Tree) hybrid machine learning algorithm. According to their experiments, the accuracy results achieved with first names are higher than the accuracy results obtained with colors and user names. In a similar study, to detect the gender, Alowibdi et al. \cite{alowibdi-14} use two more profile colors in addition to their previous study: sidebar fill color and sidebar border color. They train hybrid  classifiers using Naive Bayes and Decision trees algorithm. Based on first name (f), user name (u), colors (c) and all-combined, gender accuracies are obtained as 82\%, 70\%, 75\%, and 85\%, respectively. Furthermore, \textit{male trending} factor is calculated for each name: $m = (w_f*s_f + w_u*s_u + w_c*s_c) / (w_f + w_u + w_c)$. 
In this formula, $w$ values refer to weights whereas  $s$ values refer to sensitivity values. 
$w_f, w_u$, and $w_c$ are weights of the three gender indicators: first names, user names, and the 5 combined color characteristics. $w$ weight values correspond to the difference of accuracy from baseline: $w_f= 82\%-50\%=32\%$, $w_u= 70\%-50\%=20\%$, and $w_c= 75\%-50\%=25\%$. The sensitivity values are calculated by the fraction of numbers of a name tagged as male and female. For instance, the name ``Mary'' has high sensitivity for the female gender, almost 1.0, because almost in all cases ``Mary'' was tagged female. However, ``Pat'' has a low sensitivity because it is common among male users as well as female users. Collected male trending values are divided into 5 categories by standard deviation and variance as depicted in Table \ref{tab:maleindexgroups}. Weak female and weak male groups are considered as deceptive groups and after checking manually 5\% of the data, 24 out of 133 weak male and 11 out of 188 weak female users were determined to be deceptive.

\begin{table}
    \centering 
	\caption{Five groups depending on the male index}
	\begin{tabular}{ll}
    \hline
    Strong female & $0 \le m \le \mu - 2\sigma$ \\
    Weak female & $\mu - 2\sigma \le m \le \mu - \sigma$ \\
    Weak male & $\mu + \sigma \le m \le \mu + 2\sigma$ \\
    Strong male & $\mu + 2\sigma \le m \le 1$ \\
    Neutral & $otherwise$ \\
    \hline
	\end{tabular}
	\label{tab:maleindexgroups}
\end{table}

If the textual content of messages is used, language specific structures of the text may be beneficial for gender prediction. 
Ciot et al. \cite{Ciot-13} study gender inference from different categories of languages, more specifically, French, Indonesian, Turkish, and Japanese. 
Languages such as French are grammatically gendered and have masculine and feminine forms of each noun. This grammatical property can be used for identifying the gender
of a person. Although languages such as English have pronouns for genders and exceptional gender-oriented forms such as 'actor' and 'actress', there is not a masculine and feminine form of every word. 
On the other hand, languages such as Turkish and Indonesian are genderless and do not even have gender specific pronouns.
Authors have generated ground-truth gender information for regular (non-celebrity) Twitter users using Amazon Mechanical Turk by showing pictures of profiles whose gender could be identified.
They have conducted two types of experiments. Firstly, they have extracted k-top features (k-top words, k-top mentions, k-top diagrams, etc.) for each gender.
By using these features and SVM classifier, they are able to identify gender for Turkish, Indonesian, Japanese, and French with accuracies of 87\%, 83\%, 63\%, and 76\%, respectively.
When the grammatical properties of French language are considered, they are able to identify the gender with an accuracy of 90\%  using sius-constructions of French language.
For users who do not have sius-constructions in their tweets, SVM is able to reach 62\% accuracy, which is just a little bit over random 50\% gender guess.  

%\textit{Feature Comparison for Gender Detection.} 
The grammatical structures of languages may help gender detection as in French or harden gender detection as in Japanese.
Ciot et al. \cite{Ciot-13} mention the difficulty of gender detection for Japanese due to presence of thousands of unigrams in Japanese compared to tens or hundreds of unigrams in other languages if  k-top (words, digrams, trigrams, hashtags, mentions) features are used. On the other hand, Ito et al. \cite{ito-13} apply user attribute detection for Japanese users and use a bag of features from profiles and tweets (5,000 features and 30,000 features for profiles and tweets, respectively). Using tweet and profile features, they have predicted gender with $85.75\%$ accuracy. If they use the profile features of social network neighbors, gender detection is improved to $86.75\%$. Achieving $86.75\%$ accuracy compared to $61\%$ using k-top features \cite{Ciot-13} is a significant improvement. 
%This suggests that when k-top or n-gram features fail, bag-of-features can be evaluated.

%\textit{Language Preference Based on Gender.} 
It is important to analyze gender detection based on language preference.
Thomson and Muracvech \cite{Thomson-2010} have shown that there is a gender-preferential language in informal electronic discourse and textual content could be used to identify the author's gender. For example, females tend to use emotional words, intensive adverbs, make compliments, and use polite words. Males are likely to use quantity, make grammatical errors, and use profanity language. If the extracted features can identify the language usage, they should be helpful to detect the gender.  Deitrick et al. observe that use of emoticons such as ":)" or ":(" are indicators of female authorship \cite{Deitrick-2012}. For example, bigram "ov" happens to be in words such as love, lovely, loves and a good feature for female authorship. Using n-grams may also yield features related to proper nouns such as a city name, person name, etc. These could be related to events or people of interest by users. Although n-grams corresponding to proper nouns may be meaningful for a small set of users for a short period of time, it is hard to generalize it  for large number of users and long periods. Ciot et al. \cite{Ciot-13} identify the use of words that address male and state it as an indicator of male authorship in addition to some profanity words.  On the other hand, female users have been observed to refer to the familiar addressee, a third person, or a speaker. Based on the study on the newsgroup and email dataset analysis by Cheng et al. \cite{cheng-11}, in neutral contexts such as Reuters newsgroup, the accuracy of gender detection could be somewhat lower compared to online social networks or email data. In addition to language independent features, they include psycho-linguistic and gender-linked cues in their feature sets. In their experiments, they observed that character-based features are as good as function words (around $74-75\%$). When all of their features are used, they have obtained around $10\%$ improvement. %When pyscho-linguistic and gender-linked cues are added, gender detection becomes language-oriented (e.g., English).

\subsection{Age Detection}

Distinguishing children from adults is important in social network applications. Rather than predicting the actual age of a user, defining age groups and predicting the user's age group is possible. For this purpose,  the profile pictures of online social network users can be used to determine the ages of users. Murphy et al. \cite{murphy2012role} show that humans can predict the ages of children and distinguish  juvenile from adults. It is based on the preconscious condition of human experience using the facial features and proportions. Studies show that people can estimate ages of subjects who are in 20-54 with 2.39 years deviation \cite{Burt-1995}. Since old faces age slower than young faces, this causes age estimation to be more error-prone for old people. For example, while a 35 years old person can be estimated as 45 years old, it is very unlikely that a 5 years old kid will be estimated as 15 years old. This study deploys several survey tasks to see how humans are successful on predicting the ages. The first task is to identify whether an image belongs to an adult or a child. At least 90\% of respondents are predicted correctly. The second task is to order photos (based on age) corresponding to the ages between 6 and 16 years old of a single person. The accuracies of 167 respondents are from 74\% to 99\% for two sets of data. In this task, from the comments of the respondents, it is concluded that they use the size of shoulders and proportion of facial features to make a decision. The research also includes estimating the ages of 47 subjects whose ages are between 0 and 25. 107 survey respondents are quite accurate on predicting their ages. However, it is also observed that the prediction accuracy decreases while reaching the adolescence and it increases again after adulthood. The experience from this research can be used to determine the ages of people in pictures. Moreover, it may be possible to determine the confidence level for an estimated age.

\subsection{Geo-location Detection}
  
People use social networks to hang out, cheer a player in local sports team, or discuss local candidates in elections which all may provide useful information for localization. As not all users may provide the correct information about their locations, it is valuable to predict user location based on information other than provided in their profiles. Although locations of pictures uploaded to Flickr has been studied by improving language models \cite{Serdyukov-2009,Vanlaere-2011}, determining locations of users who do not provide images is also important especially through the contents of messages. Identifying location indicator words would be required for such analysis. 

In many studies, localization of a content is done by using the local dictionary words which are manually predefined.
Cheng et al. \cite{Cheng-2010} use a supervised learning method to identify \textit{local} words that would determine a location by manually labeling 19,178 words  and using the Backstrom's model \cite{Backstrom-2008} parameters such as the frequency strength of words. 
Chang et al. \cite{chang-12} propose unsupervised measures to evaluate the usefulness of tweet words to be used for location prediction task using bivariate Gaussian Mixture Models. The major issue is to distinguish local words from the rest. While a sports team or an attraction specific to that area might be considered as local words, generic words such as 'restaurant' and 'downtown' are useless to predict the location.
The authors propose two methods to determine local words: non-localness and geometric-localness (GL). 
In many content retrieval applications, commonly used words, stop words, such as 'the,' 'you,' or 'for'  have little significance and are treated as noise. In the location prediction task, stop words also do not provide a spatial information. In fact, they are the least spatial information providers. While local words are expected to have peak locations at a few places and low frequencies for the rest of areas, stop words are expected to be used all over the areas. For \textit{non-localness} method, measuring the usage pattern of a word to a stop word usage pattern can give a  spatial information score for that word.
After determining a set of stop words, the words having top-k least similarity to stop words are considered to be local words. For example, words such as 'you' will have high frequencies all over the places, whereas 'patriots' will have high frequency mostly in New England area. 
The \textit{geometric localness} method focuses on the geographic locations of places. Basically, a local word should have a few peak positions and usually those areas are expected to be close to each other. If the number of cities with peak word usage is low and the average distance between these cities is also low, the geometric localness measure returns a high value. Similarly, top-k words with high GL value are chosen for geo-location prediction.  They have used tweets of 5,113 Twitter data as a test set and are able to achieve 0.499 accuracy (within 100 miles) and error of 509 miles (to home locations) on average using only 250 words or less.

\subsection{Profession Detection}

Twitter users may not  provide information about their profession on their profiles. %It is important to identify certain characteristics of users in many fields such as advertising and marketing.
Wagner et al. \cite{wagner-13} propose a method to automatically infer professions (e.g., musicians, health sector workers, technicians) and personality related attributes (e.g., creative, innovative, funny) for Twitter users based on features extracted from their content, their interaction networks, attributes of their friends and their activity patterns.
The features are grouped into activity, semantic, social semantic, and linguistic style features.
The set of activity features includes 1) network-theoretic features (in- and out-degree, betweenness, etc.), 2) following, retweeting, and favoriting, and 3) diversity of activities (social / hashtag / link / temporal variety, balance and similarity). In terms of diversity of activities, authors look into whether a user uses the network for social purpose, check whether interactions of users are balanced, and determine whether two users are similar. Semantic features include bag of words, latent topics, explicit concepts, and possessives. Social semantic features are related to what other users are mentioning about specific users. Linguistic style features cover features related to the use of language.
Their method employs a comprehensive set of latent features  to perform efficient classification of users along these two dimensions (profession and personality). Their work demonstrates both a high overall accuracy in detecting profession and personality related attributes as well as highlighting the benefits and pitfalls. Ground-truth data is obtained by leveraging manualy curated \textit{Wefollow} web directories. They have generated 3 datasets: 1) \textit{Wefollow} top-500, 2) user lists that have categorizations of other users made by a user on Twitter, and 3) random test dataset from \textit{Wefollow}. 
	After evaluating a number of feature selection metrics, information gain (IG) and Pearson correlation coefficient (CC) have been used in their study. With the help of a greedy correlation-based subset feature selection, authors are able to reduce around 20K features  to 100 features per category. Using Random Forest classifier, they have achieved an accuracy of greater than 90\% for many categories for the \textit{Wefollow} top-500 dataset. On the other hand, the accuracy achieved with the random test user dataset is above 80\%. Among feature sets, linguistic style features yield consistently satisfactory results.

\subsection{Multi-Attribute Detection: Gender, Age, Profession, Interests}

In some applications, detecting different attributes of a user rather than a single attribute helps to understand the interests of those users. For example, the recommendation of a product may depend on both gender and age. Women at ages 20 and 50 may be interested in different subjects. 

Peersman et al. \cite{Peersman-2011} analyze Dutch postings on a Belgian social networking site, Netlog. They categorize users into age groups such as min16 (from 11 to 15 years old), plus16 (16 and older), plus18 (18 and older), and plus25 (25 and older). They also analyze influence of gender on min16 and plus25 groups. They have extracted a variety of basic features:  character bigrams, trigrams and tetragrams, and word unigrams, bigrams and trigrams. When SVM is used as a classifier in their experiments, token (word) features usually outperform character features. Their system achieves 88.2\% accuracy for min16 vs. plus25 classification. When gender is considered, it is noticed that the system usually misclassifies within adolescents or within adults. It is observed that the language is mostly influenced by the age not the gender.

Song et al. \cite{song-13} propose a method to predict gender and age from SNS postings. Authors argue that there are two major weaknesses of the previous methods that aim to identify user attributes. The first one is to use all postings of users while many of them have no information about users. The second problem is that each posting is considered independent of other posts while there could be relationships among postings. The authors utilize multi-instance SVM (mi-SVM) for building a classifier. In multi-instance learning, SNS postings of a person can be put in a bag where each post is considered as an instance of the user. 
Typical multi-instance learning methods assume that instances are independently and identically distributed (i.i.d). However, such assumption is not true for SNS postings of a person since postings could be related to each other (i.e., postings of a person are non-i.i.d). To benefit from relationships among postings, they build multi-instance graph (miGraph) as a graph kernel function that can reflect the relations among postings. Hence, a bag is represented as a graph where nodes are the postings in the bag and edges represent the dependencies among the instances. Then the problem is converted to a classification of a bag of instances into predefined attribute labels. A classifier $F:X \rightarrow Y$ is constructed to map a user $X$ to a label $Y$. The classifier is trained by a multi-instance SVM in which each user has a bag of instances and a label for each attribute of the user. 
To evaluate gender and age prediction, they crawled manually users and their postings from Facebook, Twitter, and me2day. Their proposed method outperforms both single-instance learning and standard mi-SVM. Their method achieves $69.2\%\pm2.7$ accuracy for the gender and $40.9\%\pm0.9$ accuracy for five age groups.

Since users do not give much information about themselves in their profile, it is difficult to find or construct ground truth data to test these types of methods. Ito et al. \cite{ito-13} propose an automatic method to create/collect ground truth data for estimating the age, gender, occupation, and interest in Twitter accounts by crawling the blogs that are related to that Twitter account. Akaike Information Criterion (AIC) is used for feature selection, and Liblinear with L2 \cite{liblinear} regularized logistic regression is used for constructing estimators. 5,000 features from profiles, and 30,000 features from tweets were collected. 86K twitter users who have blog accounts were used in datasets, and the accuracy of the proposed method was compared to  REGEXP (a labeling method based on regular expressions that evaluate words in user attributes and profiles), HUMAN \cite{Ikeda-12}, and D1000 (the method of authors with data size same as HUMAN and REGEXP) in each category (age, gender, occupation, interests). Since the proposed method has 9 to 35 times more training data (depending on the attribute) the prediction accuracy is higher than these compared approaches. 
The authors conduct experiments to determine if their method is sensitive to false information. They have performed analysis by filtering data based on equivalence of estimator results from blog articles, tweets, and blog's profile. The main idea is to observe how much filtered data could improve the accuracy. It is noted that their method is not affected negatively by false data in the training set, and filtering is not necessary for the proposed method. The last experiment was conducted to check the validity of attributes collected by neighbors of the users using the Mention relations of Twitter. Using both tweets and profile documents of a target user along with profile documents of neighbors respectively perform well for all attributes than other combinations of user/neighbor data. In general, increasing training data improves accuracy but neighbor data may be noisy if the connection of two users or the homophily (tendency of individuals to associate and bond with others) is weak. 

%\textit{Age Estimation Using Photos and Language Use.} 
%For gender detection especially based on textual features, age information could be utilized. The profile photos could also be utilized for gender detection. Although some profile photos may not correspond to the actual photo of a person (e.g., nature images, photos of relatives, etc.), they could still be used as an additional feature or to support the findings using other features. 
%The study by Murphy et al. \cite{murphy2012role} encourage using photos for age detection especially for adults. 
It is not always clear whether using multiple attributes will generate better results or not.
Peersman et al. \cite{Peersman-2011} show that 
%people are able to estimate the age group based on the photos. Particularly for adults, ages could be estimated with good accuracy. 
it is possible to separate min16 (11 to 15 years old) and plus25 (25 and older) groups with $88.2\%$ accuracy. However, gender information did not help much in their context, as they determine that language is influenced by the age rather than the gender. On the other hand, the work by Song et al. \cite{song-13} achieves $40.9\%\pm0.9$ accuracy for 5 age groups and $69.2\%\pm2.7$ accuracy for gender. These results show that it is important to identify the groups of users where the combination of age and gender could be used together.

\subsection{Discussion}

Gender detection has been studied more often than other attributes.  Features from profile, linguistics, and social neighbors are extracted for attribute prediction. 
%The profile features generally include colors used in the profile, names and images of users. Linguistic features could be language independent such as n-grams, or pyscho-linguistic or gender-related cues depending for a specific language can be extracted for classification. 
If pyscho-linguistic and gender-linked cues are used, gender detection becomes language-oriented (e.g., English). This would require developing both gender-linked and language-oriented cues for each language. The performance of such cues should be studied for different languages and comparative analysis among languages  could be used to validate this type of features. 

Features based on n-grams are frequently used for attribute detection due to its simplicity and fair successful results. Although n-grams corresponding to proper nouns may be meaningful for a small set of users for a short period of time, it is hard to generalize it  for large number of users and long periods. Moreover, in some languages such as Japanese, unigrams could be dominant and n-grams may not be effective for gender detection. Therefore, when n-gram or k-top  features do not yield satisfactory results,  bag-of-features can be evaluated.
Moreover, the grammatical features of languages that could be indicators of gender have not been studied well in the literature. Grammatical features (e.g., genderless vs grammatical gender) could benefit gender detection. 

Age could be a factor in language preference, and benefiting from age information could help gender detection. Analyzing gender based on language preference may not be an effective way since there could be difference between the language used by teenagers and adults of the same gender. Linguistic features for gender detection would be more effective for a small group of connected people rather than a diverse large group. Future research could study language preference by using both age and gender information.  

First names of users, user names, and color profiles could be combined with other linguistic and grammatical features for gender detection.
In different parts of the world, last names may also be an accurate indicator of  gender, and they could be utilized as well. In addition to color profiles, the profile photos could also be utilized for gender detection. Although some profile photos may not correspond to the actual photo of a person (e.g., nature images, photos of relatives, etc.), they could still be used as an additional feature or to support the findings using other features.

%Information from social neighbors could be aggregated for better prediction of user attributes.

The research on detecting other attributes is not as popular as gender detection. 
%Identifying actual geolocation could help to determine where the person is actually located and check whether other related information is correct. Determining professions or interests of people could be used for effective marketing. 
Future research should also focus on other attributes (e.g., education level, socio-economic status, etc.) besides gender and age.

%% file: 3user_behavior.tex
\section{User Behavior Analysis} \label{sec:userbehavior}
Behavior analysis on OSNs may help on marketing strategies, protecting privacy, security, identifying radical behaviors, and improving use of OSN for education.
In this section, we briefly look into behavior models,  deceptive behaviors, privacy behavior,   reactions to social attacks,  view-like behaviors, future behavior analysis, radicalism,  impact for higher education, and anomalies in social networks.

\subsection{Behavior Models}
Psychological personality may influence preferences for communication styles in the digital world. The messages posted on online social networks may be used to analyze user behavior. Some previous work \cite{Golbeck-11-a}, \cite{Quercia-11-a} has illustrated that people reveal their personalities through using  social media (e.g., on Facebook and Twitter), and using online social networks can lead to a high accuracy of the personality characteristics prediction by analyzing the data sets that people share online.
Adali et al. \cite{adali2012predicting} provide an analysis about to what extent behavioral features could be used to predict personality. To analyze a person's behavior towards her friends and followers on Twitter, new measures are developed based on the frequency and intensity of interactions along with the priority and reciprocity.
They have used Big Five Behavioral Model having the following traits: openness to experience, conscientiousness, extroversion, agreeableness, and neuroticism. \textit{Openness} represents curiosity, intelligence and imagination. High scorers of openness trait are sophisticated and artistic  and welcome new  experiences, views, and ideas. \textit{Conscientiousness} represents responsibility, being organized, and perseverance. People with high scorers in this group are good achievers and planners.  \textit{Extroversion} represents outgoing, amicability, and assertion. The people with high extroversion scores are energetic and friendly and inspired from social situations. \textit{Agreeableness} indicates cooperation, helpfulness, and nurturing. These people  are peace-keepers, (generally) optimistic, and have trust of others. \textit{Neuroticism} represents anxiety, insecurity, and sensitivity. Neurotic people are tense and easily influenced by negative emotions.
Authors consider actions in the following categories: (1) network bandwidth (the volume and frequency of messaging), (2) message content (e.g., links, forwarding), (3) pair behavior towards friends and followers, (4) reciprocity of actions, (5) informativeness, and (6) homophily (all features to determine user's circle). 

 Many features have reciprocity property. For example, $A\rightarrow B$ may indicate the number of messages ($x_1$) from A to B, whereas $B\rightarrow A$ indicates the number of messages ($x_2$) from B to A. The probability can be computed as $p= x_1/(x_1+x_2)$ and then the entropy can be calculated as 
\begin{equation}
H\left (x1,x2  \right ) = -p log p - \left (1-p\right )log\left (1-p\right )
\end{equation}
Entropy increases as $|x_1-x_2|$ decreases. The delay feature computes the average number of messages received after arrival of a message from a specific person. Similarly, the priority feature measures how many times a person is preferred to someone-else's message. Propagation measures the pairs of messages forwarded or shared coming from a specific user. Authors cover a good set of features for personality prediction in their paper.

A personality test, 44-question version of the Big Five inventory \cite{john-91-a}, has been conducted on 60 Twitter users. Utilizing the friendship graph of these people, around 280,000 unique pairs of individuals (who have at least sent one message) have been used for the analysis. To predict the score of personality features, ZeroR and GaussianProcess regression algorithms are used from Weka machine learning tool \cite{Hall-09-a}. The mean absolute error rate for predicting agreeableness and openness is around 11-12\%. Neuroticism, on the other hand, is much more difficult than others to predict and could be predicted  with an error rate of  around 18-19\%. The error rates for extroversion and conscientiousness error rates are in between 14\% and 16\%. It is shown that behavior of the user towards  his/her friends and followers has big insights on understanding personality of the users. They also include analysis on whether users  treat some friends similarly or differently.  The features introduced in this study can be used as strong predictors of personality.

Behavior analysis for personality prediction in OSNs is a difficult task as a person's behavior may not be consistent all the time and 
may vary depending on the context and external factors. However, the personality affects how a person uses OSNs \cite{AmichaiHamburger-2010}, and 
it may be possible to predict personality traits in a specific context.
Adali et al. \cite{adali2012predicting} observe that timing between messages, text length and propagation are good indicators for personality. 
Responses of friends and followers were effectively used to predict a person's personality. 
While large standard deviation of text length and no propagation implied high neuroticism, long text length and no propagation implied high extroversion.
The accuracy of personality prediction for neuroticism and extroversion was not at the desired level compared to openness, agreeableness, and conscientious. 
Adali et al. \cite{adali2012predicting} state that extroversion is related how the behavior of a person varies for a specific personality trait 
and neuroticism is related with the variation across traits based on Moskowitz work on flux, pulse and spin \cite{Moskowitz-2004,Moskowitz-2005}.
%The limited or no propagation may not provide enough information and additional data may need to be used for these 
%personality traits. 
%Adali et al. \cite{adali2012predicting} indicate that timing between messages are critical for prediction of personality traits. 
%Darmon et al. \cite{darmon2013predictability} propose a model for predicting whether a person
%will tweet in the future. 
%Such future prediction can be analyzed with respect to the personality traits and behavioral model of a person. 
%However, Darmon et al. do not include  interactions with other OSN users. 
%This could help to remove tweets that are unbiased (a person tweets a message regardless), and maybe accuracy of personality trait prediction could be increased.

\subsection{Deceptive Behaviors}

Understanding motives that lead OSN users to provide false information or withhold information may help to determine  the credibility of information on OSNs. Moreover, as video conferencing tools such as Google Hangout are available for free, analyzing face expressions may help to identify cases of lying.

Rather than categorizing OSN users as trustable or liar, the cases when people are likely to provide false information should be analyzed. Not all information a person provides might be correct as well as not all provided information might be false. Squicciarini and Griffin   \cite{squi-14} analyze the reasons behind  deceptive behaviors of online users by analyzing the posts of a forum. A game theoretical model is proposed based on three factors (i.e., peer pressure, potential reward, and comfort level) that affect user behavior. They  conduct an online survey on undergraduate students to have an understanding of when, why, and how users deceive. The survey includes questions on whether they would provide true or false information or withhold information about their GPA, job, age, phone, location, etc. The survey results indicate that the reasons behind providing false information or withholding data are not due to privacy concerns but because of portraying successful social character. It has been determined that users actually play a coordination game that would minimize their social stress by selecting a deception level. Withholding information and lying are two different behaviors, and deceptive behavior is influenced by inner-circle users. This might mean that rather than having a good social image online, the user environment and attitudes of close friends or people to the user may have higher influence on deceptive behaviors. For experimental evaluation, 1400 users who had at least one infraction has been selected out of 3.7 million posts of 21K users. 356 of those users had at least two infractions. The website and other comments about user activities are used to determine infractions. It is concluded that identifying deceptive users is not an easy task; moreover, deceptive users also influence other users to have deceptive behavior.

Social network sites require users' certain information in the registration process and users usually try to minimize the information they enter. This causes security and privacy issues which needs to be investigated. Griffin et al. \cite{griffin2012toward} extend their work in \cite{squi-14} on the deception model which uses game theory to detect users' willingness to release, withhold or lie about their information.  
The goal with this model is to show the complexity of information revelation in OSNs. Misrepresentation in three cases is investigated: truthful information disclosure, withhold information, and deception. Survey data is collected from users to detect typical behaviors of users' identity. The results show that the tendency of a user to lie is highly correlated with users' desire to show a good social image rather than privacy concerns.

Psychologists have shown that human behavior deviates from normal behaviors when the person lies. Bhas et al. \cite{bhas-11} focus on the eye movements to detect the deceits. In their  experiment, 132 subjects were interviewed with two types of questions: 1) questions involving basic conversations, and 2) critical questions which involves reward or punishment. These interviews are video recorded and the subjects fill out questionnaires to indicate if they lied or not.
The eye pupils are detected by using image processing techniques, and the X and Y locations (coordinates) of the pupil regions are identified to learn the behavioral features. 
In order to learn the normal behavior of a person, Bayesian model of eye movements is used in the basic conversations period. 
The rest of the interview is divided into chunks of time and each chunk is tested against the trained model by computing its log-likelihood. 
Deviations from normalcy are tested using the log likelihoods of each chunk. Deceit and non-deceit show very distinctive patterns.
In the testing part of the process on the 40 subjects,  deceit detection accuracy is 82.5\%. 

%Griffin et al. \cite{griffin2012toward} study the deceptive behaviors and categorize deception into three levels: truthful information disclosure, 
%withholding information and deception. 
%They analyze why people would like to withhold information or provide false information about themselves. 
%They realize that having good social image is more important factor than privacy for providing incorrect information. 
%Teoh et al. \cite{teoh2014gender} also indicate that privacy is not a major concern for higher education while using OSNs. 
%Deceptive behavior may also trigger deceptive behavior as deceptive behavior is more influenced by inner-circle friends.
%As eye movements may be used for deception \cite{bhas-11}, real-time deceptive behavior may be incorporated for OSNs that support video chat.
%The behavior models and personality traits could be studied to analyze trust among users \cite{adali2012predicting}. Griffin et al. \cite{griffin2012toward} suggest future work on identifying motives for deceptive behavior. The passive (number of profile views or blog visits) or active (commenting, or sharing) social 
%transactions may influence deceptive behavior. Identity revelation through secondary (triad, friend-of-friend) relationships could be analyzed. The deceptive behavior of a person may not be the same over all communities. A person may provide correct information for LinkedIn, but provide false information on Twitter. Determining the roles of communities for deceptive behavior would be interesting to analyze.

\subsection{Privacy Behavior}

Finding an effective and practical way to quantify, measure, and evaluate privacy is one of the most difficult tasks in OSNs. 
Wang et al. \cite{wang-13} propose a model of privacy information disclosure. In their framework, an actor is a social entity (e.g., person, organization, etc.) in a social network. Actor has certain characteristics that are used to describe its features, also known as attributes. Each attribute has a different impact on privacy called as Attribute Privacy Impact Factor (APIF). It reflects the sensitivity of the attribute and a large number indicates high sensitive information. The probability in their context is related to the information visibility. While probability of 0 represents an unknown value, the probability of 1 indicates known attribute value. The product of sensitivity of an attribute with its probability indicates how much private information is revealed based on that attribute. Based on this privacy model, three privacy indexes (PIDXes) are proposed based on three different privacy measurement functions \cite{wang-13}: weighted privacy index (w-PIDX), maximum privacy index (m-PIDX), and composite privacy index (c-PIDX). The weighted privacy index is actually obtained by summing up the products of sensitivities and their corresponding probabilities and then normalizing by dividing by the sum of sensitivities. The maximum privacy index returns the maximum product of any sensitivity and probability pair.  All proposed privacy indexes are normalized and mapped to values between 0 and 100. The composite weighted privacy index is weighted sum of the weighted privacy index and the maximum privacy index.

Combining a set of attributes may reveal more information than they are used individually. Wang et al. \cite{wang-13} also introduce \textit{virtual attribute}  for social network actor model in order to describe the combined attributes' behavior. In addition to individual attributes, they also study how much virtual attributes (a set of attributes) may disclose privacy. The disclosure of privacy is measured using privacy indexes mentioned earlier. If it is lower than a given threshold $T$, privacy is considered to be preserved. For evaluating privacy indexes, preliminary attributes are extracted from personal profiles in social network sites  and privacy settings. User's rating of the privacy impact of each attribute is collected with a survey. 20 attributes including name, SSN, birth date, education, etc. are selected for testing, and a privacy impact factor is assigned to each attribute. The three PIDXes are evaluated when known attributes change incrementally for different user groups. While w-PIDX is not good for privacy ranking, m-PIDX is not good for measuring privacy increment change. Tests and analysis show that composite privacy index is the best to measure privacy for social network actor model.

Balancing between data disclosure and privacy preservation is not a simple task. Different approaches are proposed to improve privacy preservation. Sayaf et al. \cite{sayaf-13} compare privacy control approaches between technological and legal frameworks. In other words, they look into whether the privacy control proposed by technical methods (e.g., methods by social software providers) are consistent with legal frameworks (i.e., laws). In terms of technical methods, they consider access control methods (ACMs) and accountability methods. These are compared with the Directive 95/46/EC of the European  Parliament and of the Council 'on the protection of individuals with regard to the processing of personal data and on the free movement of such data' as the legal framework. Privacy control is compared with respect to protection of data, linkability to the data subject (i.e., user posting the data), control of the audience, and control of context. ACMs solve privacy of the data control by offering methods that give as much control as possible over data for users. Although ACMs may be difficult to use, they provide more control on personal use exemption than laws.  However, ACMs do not provide as much control over the behavior and conduct of social network providers and third parties as legal frameworks do. To identify misconduct in social software, accountability and audit approaches verify the behavior of users by auditing the logged actions of users. These approaches are complementary to ACMs. The most important point is that there are conflicts between technical and legal frameworks and authors suggest that the most ethical framework can be achieved by integrating the two frameworks.

It is also important how information about a person is revealed as an outcome of search results.
Paradesi et al. \cite{paradesi-12} point out the problem of the privacy of social networks and provide an interface, Policy Aware Social Miner (PASM), which makes it possible for users to create policies to guide the consumers how searches about them should be executed. People may share a content within a specific context which is not harmful, but when it is viewed together with other shared content this may picture the person in a different way or may be harmful to this person. This violation can be prevented by their proposed interface. There are two sides of the interaction:  a \textit{user} who shares the data and a \textit{data consumer} such as a person preparing an insurance claim or looking for user's profile for employment. 
Data consumer may enter the user's Facebook username, keywords to search, and the purpose of the search (e.g., employment, commercial, financial, and medical). User can keep his or her privacy by restricting access to his or her profile by adding no-employment, no-commercial, and no-medical options. Another way to keep privacy is to refuse the accuracy or the ownership of information, which is called as refutation. Users may also add words for filtering. If a post has any of these words, the post will be hidden. While this interface can be effectively used  for Facebook, it can also be used for other social networks.

\subsection{Reactions to Social Attacks}

Analyzing user actions when the user's account is hacked is important for user behavior analysis.
Zangerle et al. \cite{zangerle-14} investigate the reactions of Twitter users who noticed that their accounts were hacked and actions of these users after being hacked. After manual exploration of the tweets, users are categorized into 7 classes.
\begin{itemize}
\item Class 1: Users who stated that their account was hacked and they cannot reach account credentials.
\item Class 2: Users stating that their account has been hacked and they immediately apologized for any unsolicited tweets.
\item Class 3: Users stating that their account has been hacked and they immediately apologized for any unsolicited direct messages.
\item Class 4: Users who stated that they were hacked and they moved to a new account.
\item Class 5: Users who have been hacked and stated that they now changed their password.
\item Class 6: Users who stated that they were hacked by a friend or relative, where 'hacked' refers to e.g., leaving a device unattended.
\item Class 7: Other tweets not belonging to any of the above.
\end{itemize}

In the collected ground truth, 23.36\% of all examined users state that they were hacked in simple terms while 27.55\% of them apologize for inconvenience because of unsolicited tweets and direct messages. The Support Vector Machine (SVM)  is used to classify the tweets. The overall accuracy of the experiment is 78.25\%. The analysis of errors shows that most of the classification errors are between the class 2 and class 5. Furthermore, overlap between classes 2 and 3 which includes apologizing for tweets and direct messages causes classification error.

\subsection{Future Behavior Analysis}
Analyzing past behavior of users for predicting their future behavior may be helpful to determine responsive users of the social media, and this information can be used for marketing purposes as well as for understanding how people interact with the social media. Darmon et al. \cite{darmon2013predictability} aim to capture the behavior of users on social media by modeling them as computational units processing information. The main purpose is to predict whether a person will tweet in a specific time interval or not.
An individual using a social media service is treated as a computational agent. A social media user will behave based on the information obtained from surroundings after evaluating them with respect to the internal states of the person. User behavior is modeled as binary time series. From the tweet times, a binary time series is created in such a way that $X_i$ is 1 if the user tweeted at least once in the time interval $\Delta t$, and 0 otherwise. In this manner, a user's behavior is encoded as a sequence. Then, one-step ahead prediction is performed to predict $X_i$ based on the past behavior of the user as a finite history $X^{i-1}_ {i-L} = (X_{i-L}, ... , X_{i-2},X_{i-1})$ of length L. Two types of approximate models are used to predict user behavior: 1) causal state models inspired from computational mechanics and 2) echo state network, which is a type of neural network that can address complex dynamic systems by training only simpler weights and using randomly selected and fixed weights to drive the recurrent activity. Causal state modeling is based on a simple model and its structure is enhanced as needed whereas echo state networks consider behavior as a result of complex internal states with intricate relationships which are simplified later. For evaluating and comparing these models, 12,043 Twitter users' statuses  over a 49-day period were collected. Then, only the top 3,000 most active users over this period are selected. For these user's tweets  only the ones between 7 AM and 10 PM (EST) were processed. After that, the status of each user was transformed into a binary time series with ten minute intervals. 49 days of user activity were partitioned, chronologically, into a 45-day training set and a 4-day testing set. Two methods perform very similarly on a large proportion of the users although both methods are naturally different from each other. When compared to a baseline classifier based on the proportion of ten-minute windows with tweets over 49-day period, the performance of these models is especially significant for users whose tweet rate is above 0.2.

\subsection{Detecting Radicalism}

Online social networks can be used to identify radicalization using the content and geolocation of tweets of a user. Mazumder et al. \cite{Mazumder-13} state that online social networks provide large amount of information about human feelings and emotion about people, places, events, and political activities. As part of the Minerva project \cite{Mazumder-13} at Arizona State University, information has been collected to understand political activities of Indonesia. Based on these data, a heat map is generated for the radical activities in provinces of Indonesia by computing {\em radicalization index} and {\em location index} of each Twitter user from Indonesia. Rather than considering tweets individually, all user data are analyzed. A {\em degree of radicalism} is assigned to each user based on the content of his or her tweets. Since Twitter API2 provides geolocation of tweets, it is possible to track the location of a user based on his or her tweets. Since tweets for the same user might have been posted in different provinces, a probability distribution estimate is generated per user. Results show high accuracy in detecting radicalism in users' tweets. The higher value in the radicalism index, the higher radicalism there is in the province.

If the user connects to a remote system and posts message through such a system, Twitter API may not return the actual location of the user. 
In such cases, the localization index should be supported with geolocation prediction as mentioned in Section 2.3 using the content of messages as suggested in \cite{chang-12} and \cite{Cheng-2010}.

\subsection{Impact for Higher Education}
The use of social network sites (SNS) in higher education has been studied to analyze its effects on the performance of students.
Certain findings suggest that social media only distracts students and affects their academic study negatively, while other findings suggest that it  helps students to communicate with teachers. Teoh et al. \cite{teoh2014gender} summarize the findings in the literature and study SNS usage by different genders in higher education. Their main goal is to point out the factors that have a positive effect on men and women in higher education.
Research in SNS is categorized into 1) privacy and disclosure, 2) SNS structure and social capital, 3) identity, personality, and behavior, 4) applied SNS, and 5) gender and SNS.
	 	 	 	
\textit{Privacy and Disclosure.} Teachers' self disclosure on SNS seems to cause students a high level of motivation, effective learning, and positive classroom climate. Men and women seem to have different levels of self-disclosure on SNS. According to the observations, privacy concern is not a big issue for SNS in higher education. However, inappropriate disclosure of information of teachers on SNS may cause loss of respect by the students.

\textit{SNS Structure and Social Capital.} According to Coleman \cite{coleman1988social}, social capital is “the resources accumulated through the relationships among people.” In SNS, the resource is generally 'friends.' People tend to follow others with various closeness degree. According to the early research, it was believed that users tend to connect with people other than those whom they interact regularly. However, it was found that this is not always the case. McMillan and Morrison~\cite{mcMillan_and_morrison} found that Facebook users for instance, prefer to be friends with those who are within their daily life circles. 

\textit{Identity, Personality, and Behavior.} Some research studies indicate that SNSs are helpful in terms of users' self-esteem and life satisfaction improvement. When users are supported by others, the behavior seems to spread more rapidly. It is also found that the personality of users also has an impact on how they use SNS.

\textit{Applied SNS in Higher Education.} There is no consensus whether SNSs improve students' academic performance or not. While some suggest that it has positive effects on student engagement,  some suggest that SNSs such as Facebook have negative effects.

\textit{Gender and SNS.} Male students seem to prefer more involvement of faculty on Facebook than female students. Female students prefer to use technology to create relationships rather than to use it for learning.
 	 	 	
The framework that is used by this research is based on the Push Pull Mooring theory (PPM) \cite{lee1966theory}\cite{moon1995paradigms}. 
As the name implies, the theory has three variables: \textit{Push} variable covers the factors that make people migrate from a location such as high crime rate. \textit{Pull} variable covers the factors that make people migrate to a location such as better education. \textit{Mooring} variable covers the factors that make people stay at their location such as family obligations. The PPM was adopted by other researchers to explain behaviors of SNS. Since this study is based on a previously conducted research in Malaysia, the same PPM factors are used. E-learning perception is the push factor. Convenience, social influence, academic reasons, ease of use, and social networking are the pull factors. Barriers are the mooring factors.

The data gathered from both Malaysia and Australia are used to see if the previously mentioned factors had an impact on different genders. The data shows significant differences among the genders in higher education in all of the factors except for the barriers, electronic learning perception, and convenience. Male respondents perceive more teaching and learning benefits in using SNSs than female respondents. Male respondents experience a higher degree of social influence to get involved in SNSs than female respondents. Male respondents find academic reasons to be an important factor for SNS use as opposed to the female respondents. Male respondents think of the ease of SNS for e-learning to be an important factor.

The study shows how different genders approach the idea of higher education in SNS. The data shows that there are significant differences between  genders. Male respondents generally show higher means for perceived teaching learning benefits, social influence, academic reasons, ease of use, and social networking. The pull factors of the PPM model are social influence, academic reasons, ease of use and social networking. These factors can be indicators of what the academics can focus on in order to influence the success of SNS in the classroom.

\subsection{Anomalies}

Anomalies in online social networks may be indicators of malicious individuals such as spammers, sexual predators, and online fraudsters. The social networks for anomaly detection can be studied as static vs. dynamic, labeled vs. unlabeled, and local vs. global. Savage et al. \cite{savage-14} study computational methods for recognizing anomalies in online social networks. Social networks are considered dynamic as new users join or leave as well as the connections between users change \cite{savage-14}. If the information exchange (e.g., size, content, time) between nodes is included, the network is considered as labeled, otherwise it is considered as unlabeled. Dynamic structure of a social network can be reduced to static structure. For example, rather than maintaining data about each information exchange, the summary of exchange can be used between nodes. If the anomaly is with respect to the close neighbors, anomaly is considered to have local context. If a group of nodes are acting together and manipulating the network, the anomaly can be considered to have global context. 

The anomaly detection analysis is recommended in two phases: 1) detection of feature space and 2) analyzing anomalies in this space. These phases are further split into five steps: 1) identify the smallest unit (sub-graph) affected, 2) identify how this unit differs from the rest, 3) determine the context when the anomalies are expected to happen, 4) extract feature space for anomaly detection, and 5) apply anomaly detection methods in this feature space.

\subsection{Discussion}

Online social networks have influenced our daily lives tremendously regardless of their purposes (e.g., professional or friend networks such as LinkedIn and Facebook or content sharing networks such as YouTube and Flickr). Understanding how users behave when they connect to these sites is very crucial for a number of reasons. From a seller's perspective,  user behavior analysis may help improve the performance of marketing systems, lead to a better site design and advertisement placement policies.  From social studies perspective, accurate models of user behavior in OSNs are crucial in understanding user demeanor and how users react to events in social environments. Despite some promising results, there is still some work to be done for behavior analysis.

%Behavior models can be used to determine one's psychological state. This can help either take pre-cautious steps in order to prevent an anomaly or take pre-emptive steps in order to provide better service to users. People mostly share what reflects their current psychological state on the internet which makes it easier to perform behavior detection and analysis. Therefore, new measures along with methods are developed in order to provide such functionality to systems which perform human-behavior analysis.

%Even though it is not an easy task,the future of the behavior analysis research on online social network can be contemplated with some ideas such as (i) studying the role of communities for deceptive behavior with respect to confined information and deception, (ii) studying pre-existing motives of deceptive behaviors, (iii) data traffic-flow analysis of the  malicious users (e.g., hackers),  etc.

%--------

Personality trait analysis may not be performed effectively in case of limited or no propagation and may thus require additional data for predicting traits. Since personality trait analysis may benefit from timing between messages \cite{adali2012predicting}, a tweeting behavior model of a person as proposed by 
Darmon et al. \cite{darmon2013predictability} could enhance the prediction of personality traits. Use of biased or reactive messaging/tweeting could be a better indicator of a personality trait, and removing tweets that are sent regardless could help improve the accuracy of personality trait detection.

%The limited or no propagation may not provide enough information and additional data may need to be used for these personality traits.

%Adali et al. \cite{adali2012predicting} indicate that timing between messages are critical for prediction of personality traits. A tweeting behavior model of a person in future such as the one developed by 
%Darmon et al. \cite{darmon2013predictability} could be used for prediction of personality traits. 
%%propose a model for predicting whether a person
%%will tweet in the future. 
%Such future prediction can be analyzed with respect to the personality traits and behavioral model of a person. 
%However, Darmon et al. \cite{darmon2013predictability} do not include  interactions with other OSN users. 
%This could help to remove tweets that are unbiased (a person tweets a message regardless), and maybe accuracy of personality trait prediction could be increased.

%The behavior models and personality traits could be studied to analyze trust among users \cite{adali2012predicting}. %Griffin et al. \cite{griffin2012toward} suggest future work on identifying motives for deceptive behavior. 

It is important to distinguish between deception and unintentional mistakes. While identifying incorrect information is important, it is also crucial not to label people based on deception. There are many reasons for incorrect information sharing. For example, after selecting an item from a combo box, scrolling down/up with a mouse may change the selected item. Inconsistent information may be an indicator of an unintentional mistake. Moreover, it is also likely that a person may not necessarily update his or her information and inconsistencies may happen. Besides identifying false information, it is important to analyze the motives for intentional deceptive behavior. 
%Identifying the motives for deceptive behavior rather than identifying deceptive information could help research study the relevant people and their attributes. 
Griffin et al. \cite{griffin2012toward} list some future work for analyzing deception.
The passive (number of profile views or blog visits) or active (commenting, or sharing) social 
transactions may influence deceptive behavior. Identity revelation through secondary (triad, friend-of-friend) relationships could be analyzed. The deceptive behavior of a person may not be the same over all communities. A person may provide correct information for LinkedIn, but provide false information on Twitter. Determining the roles of communities for deceptive behavior would be interesting to analyze.

%It is important to determine the attributes that may carry false information and reasons for those specific attributes. Griffin et al. \cite{griffin2012toward,squi-14} state that the user will not benefit from incorrect information about his or her age, gender, personal website, etc. However, users may hide or lie about dating status, which could be hard to determine the actual status. 
%It is important to know what attributes are preferred for lying or withholding information and the reasons behind these behaviors.
%For example, a person may not provide SSN or date of birth for privacy reasons, 
%but a person may lie about his or her GPA or voluntary work to have a good social image. 
%There are also malicious people who have bad intentions and could provide incorrect information about their gender or age. 

Categorizing people based on information sharing should not be limited to sharing regarding true information. 
%It may be helpful to categorize users based on how they share information. 
We should consider groups of people who share false information as well as true information. 
To cover deceptive information sharing, two more categories (i. people who lie for representing a good social image and ii. people who lie for malicious intentions) could be added to three categories of information sharing mentioned in  \cite{wang-14}:
 privacy fundamentalist (unwilling to share), 
pragmatic majority (willing to share with privacy control), and marginally concerned (willing to share).
%These may provide user groups who are willing to provide true information. For deceptive behavior, 
%we may add two more categories:
%people who lie for representing a good social image and people who lie for malicious intentions. 
Then it would be good if these five groups of people could be linked to behavioral models. For example, it would be nice to determine whether there is a relationship between deception and neuroticism or not.
%Peer pressure and establishing a good social image are main motives for revealing private, detailed information as well as lying \cite{griffin2012toward}.

Privacy may affect the behavior of information disclosure on OSNs. 
Effective methods are required for maintaining privacy. The legal (i.e., laws) and technical (e.g., access control methods) frameworks should be merged for maintaining user privacy. 
%When information is disclosed, a quantitative measure is needed to determine whether privacy has been violated. 
%This can be achieved by assigning sensitivity to attributes of users as in \cite{wang-14}.
%Paradesi et al. \cite{paradesi-12} study how user's privacy may be maintained and propose a policy aware social miner. The user may restrict the results of searches for specific categories and provide refutation for some search results.
%Their current system is keyword-oriented where the data subject identifies keywords and provides restrictions or refutations based on those keywords.
Automatic restriction (for returning results) and refutation methods are needed based on prior restrictions and refutations. 
%A user may not be able to follow on updates. 
It is important to notify a data consumer after his or her query results might have been changed after a while without requiring the user to perform the query again. It may be critical to inform the users about the best up-to-date valid information.
%The method could be more active and continuously search documents about a person and inform the user about what restrictions or refutations should be applied.
%Those documents could be clustered and automatic restrictions and refutations with the approval of the data subject can be implemented as future work.
%Another issue is that the proposed system is pull-based system where the data consumer queries the system and then uses the results. 
%As in push-based systems, the data consumer could be made aware
%of new results, restrictions, or refutations as they are available.

%In addition to study of Zangerle et al. \cite{zangerle-14},which is the immediate reaction of users of hacked accounts, a f

There are also other areas where future work could be performed. Real-time deceptive behavior analysis may be incorporated for OSNs that support video chat.
%Future work may also include how hacked users approach using OSNs after being hacked. 
Researchers may study whether hacked users abstain from OSNs, continue to use as before, withhold information, or lie about credentials. It may be interesting to 
analyze hacked user behavior with respect to the personality traits.
In addition, gender issue has not been studied along with the personality traits. 
%Based on literature survey and their study, Teoh et al. \cite{teoh2014gender} indicate that male and female use OSNs for different purposes for higher education. While male user utilize OSNs for academic reasons, building relationships looks to be important for female students. There are gender specific differences how female and male users benefit from OSNs. 
Gender-based personality analysis could be studied to see how gender affects behavior for a specific personality trait.

%Mazumder et al. \cite{Mazumder-13} study radicalization behavior using location index based on geolocation information provided by Twitter API2. 

%Finding users that do not behave as consistent as their network may help to find malicious users \cite{savage-14}. %Views and likes/dislikes actions could be used to predict the personality traits of a person. The studies may analyze how person's behavior (views, likes, dislikes) is affected by his or her community.

%% file: 4mental_models.tex
\section{Mental Models} \label{sec:mentalmodels}
 
Understanding mental models of users is important to analyze what types of languages they use and how their opinions change. Developing mental models may help to imitate user behavior, and especially those models can be used to improve the user interface for better communication and protect users at OSNs. A change in a mental model can be detected with analyses of the word change of its context. People change word usage in their conversation when they are communicating inside a group of OSN. There are also differences in word usage in different geographical locations. Political preferences of media outlets can be extracted by analyzing their followers’ twits. OSNs analysis metrics can be used to investigate gender based mentality for collaboration in publications.

\subsection{Security Mental Models}

Reasoning patterns of the mental security models of human beings can lead to systematic behaviors on different tasks.
A computer system which has the ability to mimic human activity would be useful to protect people from many attacks since they are already dependent on users.
In order to generate better interfaces for effective communication with the users, mental models in computer security have been adapted to the security models of users’. It was observed that users can merely look for the attacks or apply security tools that they are already aware of.
Blythe et al. \cite{blythe2012implementing} show that it is possible to reproduce security behavior observed from non-experts.
Mental models are analyzed to guide home computer users to make a decision on which expert security to adapt.
Authors implement eight mental models which were identified by Wash et al. \cite{Wash-11} for 'viruses' and 'hackers'. Namely, these mental models are buggy, mischief, crime, burglar, vandal, and big-fish models. 
In order to select mental models, each alternative course of action is simulated to decide on models which have mild security consequences.
It is shown that the results of the simulation performed by the authors matched the results of the survey by Wash \cite{Wash-11}.
Keeping patches up to date, cautious website visits, making regular backups, and using anti-virus programs are some of the considered security behaviors. 
As a conclusion, the authors find that, in their implementations, as a result of several tests performed, “mischief” and “vandal” models are found to be helpful for the protection, and “crime” and “burglar” models are found to be helpful in prevention of access control. These results were obtained from responses to questions about (1) using anti-virus software, (2) exercising care in which website to visit, (3) making regular backups and  (4) keeping patches up to date.

\subsection{Language Usage Analysis}

Our social identity  has an important effect on the way of language usage. Different styles  of language usage can be strongly associated with social or cultural groups. 
There are some studies to understand social identity of users by using language as a proxy for social behavior.

Social network analysis can be performed for analyzing user behavior inside and outside of a community. Tamburrini et al. \cite{tamburrini2015twitter} show that Twitter users  change their language depending on whether they communicate inside the group or outside the group. They also examine the correlation between the level of variation on the language usage of a community and how strongly the community is linked.
$189K$ Twitter users' data were collected between 2007 to 2009, yielding 75 millions of messages without retweets, that were copies of other messages. The data is divided based on a modularity maximization to 414 groups then filtered to 42 groups having more than 250 users and then to 24 having only English messages. Euclidean and Jaccard measures are used to determine distances of inner and outer messages based on the following frequencies: 1) word-usage, 2) word-ending, and 3) apostrophe usage. Variety of distances between the internal and external word usages are computed. A bootstrap by resampling (with replacement) new random pairs of collections of messages from the union of the original internal and external collections is created. And this process is repeated for 1000 times for each group, and the p-value (proportion of resampled collections whose linguistic distances are above internal and external collections) is calculated to make sure that these differences in word usage did not happen by random chance. The result suggests that users change their words and word-ending usage according to whether they are messaging other members of the group or not $(p \leq 0.001)$. On the other hand, for distance between internal and external apostrophe usages,  17 of the 24 groups are found to be significant $(p \leq 0.05)$.

\subsection{Opinion and Interest Analysis}

Golbeck et al. \cite{golbeck-14} propose a methodology to estimate political preferences of media/news outlets or similar think-tank groups based on their followers. In contrast to other studies, this work does not use the news outlets' own content but only their followers'. This study is done for the 2008 President election in USA.  All the followers of the congress members are collected, and data analysis is done based on these followers. The methodology has 3 steps. First for the likelihood of liberal or democratic estimation, two well-known scores, Americans for Democratic Actions (ADA) 2009 and DW Nominate scores, are calculated for a seed group. Next, the scores of the seed group are mapped onto their followers, and P-scores are calculated to see how actually they are correlated. 
Assumption on that is if someone follows a congress who is liberal, he/she is also liberal. Lastly, the scores found by the second step are mapped to media outlets (i.e., how much of the media outlets are followed by these liberal and conservative people?). This is measured by a sampling method. Followers' scores are validated by two steps.  Only the users that have the specific tweet words as hash tags such as \#voteforRomney, \#Obama2013 are filtered. This ensures that only the followers who clearly express their ideas will be examined. Moreover, these filtered 4K followers are manually tagged by the authors of the paper, whether they are supporters of Obama or Romney. The distribution of scores for followers for Obama and Romney are found to be correlated with the score of ADA and DW Nominate Scores.

Creating and sharing content on the Internet and providing opinions on the content has been simplified with a simple click. Kohli et al. \cite{kohli2014modeling} use data obtained from YouTube and certain social media to analyze some aspects of human behavior. They have collected the data set through 30 undergraduate students. The students are divided into groups of 3 and each group is asked to select 20 songs and provide their names, views, likes, and dislikes. Out of 200 song details that were collected, 130 of them are selected after pre-processing such as removing repetitions, etc. They have conducted two types of experiments: 
finding possible relationships among attributes using a machine learning techniques and finding patterns of human behaviors.
 They have observed the following results. In 119 songs, the number of 'likes' and 'dislikes' received is $0-1\%$ of the number of views for each song. The remaining 11 songs received more than $1\%$ of the views as 'likes' and 'dislikes'. The opinions for 119 songs are just from $0-1\%$ of total views. Only the remaining 11 songs got more than $1\%$ of user views. This indicates that less than $1\%$ of the population would consider sharing their opinions about what they view on the Internet. This could potentially show the lack of leadership qualities, innovative nature, and opinion expressing abilities of the users on the Internet. The dislike distribution of the songs are as follows: 21 had $0-5\%$, 62 had $5-10\%$, 28 had $10-15\%$, and 19 had $15+\%$ user dislikes. This could mean that the popularity does not mean widespread likability. The relationships between likes and views and between opinions and views are complex and not always related. The relationship between dislikes and views, however, seems to have a relationship but with some error. The paper has findings such as a) popular things/ideas do not always get accepted by the entire audience and b) only a small portion of the population tend to express their feelings and thoughts publicly.

Both textual contents and spatial features of Twitter data give valuable information about Twitter users. Zheng et al. \cite{zheng-14} combine both types of features for a better user characterization. By using (lat,lon) information, events or topics of local interests can be discovered and how a user's topical interest changes by regions can be examined. For this purpose, they use a Bayesian latent topic model. For each user, a topic distribution is drawn by Latent Topic
Model (LDA) \cite{blei-2003latent}. For each word in each tweet, latent topic is drawn on a word and spatial coordinate. Topics are quantified for their globalness based on popularity on diverse geological locations. After determining globalness of a topic, the users' interests on global topics are also quantified. For this experiment, around 620K tweet data having spatial information is collected from July $12^{th}$ to October $12^{th}$ for 9,369 users. Data perplexity is calculated to evaluate the model quality and the likelihood of test data with a trained model. After calculating user inference to see how well the combination of contents and spatial information is, the accuracy of location prediction is investigated. Using the proposed methodology, it is possible to determine the specific parts of the world having interest on global topics and  regions following local topics. For example, it is observed that users in New York, London, Texas have more interest in global topics than users in Western Africa and East India. Similarly it is possible to identify the topics that attract global attention such as London Olympics. Some outlier local news that had globalness is detected (e.g., Denver shooting). It is also noted that user having interest in global topics have distant neighbors on the social network and more mobility with respect to users interested in local topics.

Online social networks may be used to analyze changing interest or opinions of users. Grimaudo et al. \cite{grimaudo2014tucan} compare a user's Twitter data posted in different times to analyze the change of topic interest of a user, and in addition, compare that user with other users at different times to see if any correlation exists by providing results in graphs. After Twitter data collection, preprocessing is done to filter out stop words, HTML links, tag entities, and Twitter mentions. Then stemming, lemmatization, and anthology based lexicon generalization are performed. Each user's interest is represented as a bag of word model with Term-Frequency-Inverse Document Frequency and cosine similarity measure is used to identify correlation. 700 randomly selected and 28 well-known public figures are selected among politicians and news media. Only 300 users who have posted more than twice a week is included in the test. 20 users posted more than 400 tweets in a day. The window size (7/14/30/etc. days) needs to be adjusted when comparing users' tweets. Topic trend of the users from its own messages, or results of pairwise comparison of users are depicted in color graphs where color represents the normalized similarity. As an example of a single user analysis, it is observed that it is possible to detect that the user changed its social partner or dating partner from the change in the messages he or she used. It is also possible to analyze the correlation of topic interests. For example, Barrack Obama and White House tweets correlate before election. Furthermore, how this correlation changes can be examined based on the events and period of times: educational cost cut season, US immigration laws reformation, etc.

\subsection{Gender Impact on Collaborations}

Social network analysis can also be applied for academia to observe the gender impact on scholarly activities.
Ozel \cite{ozel2012link} investigates the gender impact on publication productivity, participation, presence and contribution by analyzing the data collected from social science publications in Turkey. Production and co-authorship is calculated by Lotka's Law \cite{lotka-1926}. If a paper has 3 male and 1 female authors, it is considered as having 3 male and 1 female presence. Participation of gender is 1 if there is at least one author for the corresponding gender. So, for a paper having 3 male and 1 female authors, both gender participations are 1. Contribution is the ratio of a gender to the total number of authors. For the previous example, the male contribution is $3/4$, whereas the female contribution is $1/4$. Furthermore, 3 criteria on authors' network on the social network are examined: 1) \textit{centrality}: the relative number of citations of an author with respect to citations of others in a community,
 2) \textit{clique count}: as an indicator of three or more authors having connection to each other but relatively less connection to other groups, and  3) \textit{clustering coefficient}.
7835 papers and 6738 author data are collected from ULAKBIM \cite{ulakbim}. Gender coding is done manually based on the names of authors. $38\%$ of authors are female, and $62\%$ of authors are male in the dataset. Results show that females co-author $62\%$ of papers whereas males co-author $55\%$ of papers. 
 
Overall the participation presence and contribution results show that males are more active than female in collaborations. Productivity calculation shows that males are slightly more productive than females. And statistical significance test for productivity which is provided by Wilks' MANOVA test suggests that there is significant gender difference at productivity. While there is not much significant gender difference on centrality and clustering coefficients, there is high difference in clique count and betweenness centrality.

\subsection{Discussion}

Proposed approaches for mimicking mental behavior models are too generic, mostly related to computer usage. For OSNs, more specific mental behavior models should be investigated.  Such as if a friend request comes in OSNs, what do users think and how do they decide to approve or disapprove? Does gender take a role on this decision (e.g., females tend to accept females, males tend to accept females and males as friends)? How does a social network user decide for sharing an item to certain people in his or her friend list instead of making it public? 
What makes a user to decide whether a link in OSNs is safe to click?

Alternative to manual validation of twitter followers for political tendency, retweets could be analyzed. A user retweet from that congress member is a stronger indication of the same political tendency than just following the congress member. This research may also form bias towards some media outlets or think tank organizations since a small group of actual users who follow a congress member contributes to the score of these media outlets and organizations. For example, the New York Times has over 2.5M followers in 2014 but average percentage of congress followers among these is only 1.4\% (98.6\% of users are ignored due to unavailable scores). Context analysis could also be used to give a score for political tendency besides following information on congress members \cite{abdullah-fisher-2016}.

While investigating the localness and globalness of a topic, it would be good to determine if there are connections between topics and regions. For example, news about a cop beating a person should have low globalness score, but its globalness score may increase if the beaten person is a foreign national. Furthermore, it is difficult to assess globalness and localness of topics objectively. External factors (e.g., TV coverage) that contribute to globalness should be considered. 
  
For user twitter time-line analysis, determining the interval is critical. 7-day inference may give some change in the interest of a user while 21-day inference may suggest almost no change. User interest change could be used for marketing (e.g., health-care, mortgages, etc.). Such analysis could be applied for security and forensics after attacks. For longer term analyses two users or groups of users can be compared as well. How do two groups of users think in some matters in the last 1-3 years? Is there any change in their approaches on these matters?

%% file: 5user_categorization.tex
\section{User Categorization} \label{sec:usercategorization}

There are many different types of users benefiting from as well as abusing online social network sites. It is important to distinguish real innocent users from the rest.
Propagating spams through the social media such as Twitter and Facebook or through websites such as Craigslist is very common. Even though there are some analysis tools and filtering techniques, most of the time it is not enough to prevent such spams. Thus, it is crucial to find the spam messages and spammers by creating features such as shown in Tables \ref{spammers_features}-\ref{spam_message_features} and using machine learning techniques \cite{stringhini-10,wang-14}. 
Creating honeypots and fake accounts may also be used to attract the spammers and then to find the creators of these accounts (spammers) \cite{stringhini-10,yang-14}. Developing more efficient social honeypots \cite{yang-14} helps to understand the behavior of social spammers. It is also important to analyze how spam messages affect trending topics \cite{stafford-13} and to see if there is a relationship between spammers and financial \& cultural features of a society \cite{garg-13}. In this section, we briefly look into categorization of spammers, fake users, bot accounts, and social network actors.

\begin{table}[htbp]
\centering
\caption{Features used to detect Spam Messages in Machine Learning techniques, \textit{Stringhini et al \cite{stringhini-10} }}
\label{spammers_features}
\begin{tabular}{|l|p{5cm}|}
\hline
\textbf{Feature Name} & \textbf{Explanation}                                                            \\ \hline
FF ratio              & The number of friend requests sent compared to the number of friends \\ \hline
URL ratio             & The presence of URLs in user account's logged messages                                 \\ \hline
Message similarity    & The similarity leverage among the messages sent by a particular user            \\ \hline
Friend choice         & Possibility of a list of names used by a profile to add friends                      \\ \hline
\#of sent messages    & The number of messages sent by a profile                                       \\ \hline
\# of friends         & The number of friends a profile has                                             \\ \hline
\end{tabular}
\end{table} 

\begin{table}[htbp]
\centering
\caption{Features used to detect Spam Messages in Machine Learning techniques, \textit{Stafford et al \cite{stafford-13} }}
\label{spam_message_features}
\begin{tabular}{|c|}
\hline
\textbf{Feature Name}                                                      \\ \hline
rank of topic              \\ \hline
urls per word                                           \\ \hline
total number of characters              \\ \hline
number of total characters                            \\ \hline
\ number of urls                                     \\ \hline
\ number of hash tags                                                       \\ \hline
\ number of mentions                                            \\ \hline
\ number of retweets                                           \\ \hline
\ whether the message is a reply or not                                               \\ \hline
\end{tabular}
\end{table}

\subsection{Detecting Spammers}

Spammers always look for ways to reach new victims with their unsolicited messages. Based on a market survey in 2008, 83\% of the users of social networks have received at least one unwanted friend request or message \cite{harris-08}.

Honeypots (or honey profiles) are decoy machines/servers placed within the network used to bait and alert on attackers. Stringhini et al. \cite{stringhini-10} focus on analyzing how spammers accomplish their desires on social networks. In total of 900 honey profiles deployed on 3 different social networks (in order to collect the spamming activities), the contacts and messages that are received by the honey profiles are recorded. The collected data is analyzed in order to identify anomalous behavior of users who contacted these honey profiles. In particular, some machine learning techniques are used to classify spammers and legitimate users. Six features, as shown in Table \ref{spammers_features}, are developed in order to detect whether a given profile is a spammer or not. The results demonstrate that identifying the accounts used by spammers is possible. In a real case work, after collaboration with Twitter, 15,857 spam profiles are detected and deleted.

Similarly, in another work, Yang et al. \cite{yang-14} aim development of more efficient social honeypots as well as by doing that achieving new understanding to contend against social spammers. Their intention is to formulate the priority of the active crawling/sampling of more feasible spammer accounts from Twitter, instead of aimlessly crawling all possible (or random sampling) data from Twitter. First of all, several honeypots, trained by the various well defined behavioral patterns such as Follow Behavior (Follow), Tweet Behavior (Tweet), and Application Usage (App) are placed to harvest the spammers on Twitter. Within five months, 1512 unique accounts were collected by these honeypots.  1077 unique accounts followed at least one of the honeypots, and 440 unique accounts posted at least one mention (@...) to one of the honeypots. Based on the results of the data collected by honeypots, two new sampling approaches are developed: hash tag sampler and friend sampler. In the hash tag sampler, the spammers aim to follow the accounts which post key works of the spammer's interest. In the friend sampler, spammers aim to choose the distinguished accounts' followers. As a result, more than 17,000 spam accounts are crawled within a short time (i.e., 6 spam accounts in every 10 harvested accounts).

Wang et al. \cite{wang-14} propose a framework, called SPADE, to detect spams. SPADE is an analysis and detection framework which can be used in all social networks. It brings a combined spam detection across different social networks. This framework offers several benefits. One of the benefits is that a spam detected in one social network can be detected in other social networks. Another benefit is that the detection can be improved by the cross domain classification. It can also be modified to integrate other filtering techniques such as blacklisting. In order to allow integration of multiple social networks, they propose a uniform schema model. In this model, they include three basic models: profile, message, and web page models. For each model, they use commonly used attributes. For instance, the profile model has 74 attributes whereas the message model has 15 attributes. However, the attribute count of the web page model may vary depending on the data carried by the HTTP session headers for each session. They use 4 different datasets in this work: 1.8 million MySpace profiles, over 2.4 million tweet data from profiles in Twitter, TREC dataset which consists more than 75,000 email addresses, and Web Spamm Corpus and WebBase dataset which has nearly 350,000 spam web pages. They use over 40 different types of machine learning algorithms in Weka software package to build classifiers for each model. As a result, over 0.92 F-measure and 91\% detection accuracy on web page model (in cross domain classification), 0.87 F-measure on user profile model, and 0.89 F-measure on message models are achieved by SPADE.

Stafford et al. \cite{stafford-13} investigate influence of spam messages on trending topics. In other words, they analyze whether the spam messages can drive trending topics or not. 9 million tweets across 800 distinct trending topics are gathered and around 1500 tweets are manually labeled as spam (1453) and not spam (42). Naive Bayes classifier is used to predict spam messages by using 10 attributes as shown in Table \ref{spam_message_features}. Although 90\% of the instances are classified correctly and the micro-F1 measured is obtained as 0.929, the spam detection accuracy is not outstanding due to the low macro-F1 measure which is 0.596. Chi-squared goodness of fit test is used to analyze spam variance among topics. After applying statistical significance test, it is suggested that the probability of spam being represented equally across topics was infinitesimally small, which concludes that spammers do not drive trending topics. 

Social bookmarking (SBM) enables users to store links to web documents in a centralized system. It is also possible for users to interact with this service such as commenting and tagging for searching and managing bookmarks. However, SBM has been a target for the spammers since it makes it considerably simple to modify the hyperlinks, comments, etc. It is aimed by spammers to modify the entries of SEO (Search Engine Optimization) for webpages in order to assign higher or lower values in terms of ranking for the targeted webpages. Sakaura et al. \cite{sakakura-12} propose a scheme for the detection of such attacks. The main idea is to cluster SBM accounts by the similarities of the reliable bookmarks. These bookmarks are those that trusted accounts are linked to. In other words, two SBM accounts would be considered to belong to the same cluster if they have their bookmarks linked to almost the same websites or resources. The reason for using bookmarking similarity for clustering is that the spammer-controlled accounts would most likely to point to the same websites and resources. Therefore, this system detects this type of accounts to be Intensive Bookmarking using Multiple Accounts (IBMA) spammers, which correspond to accounts that aim to increase ranking of certain pages.
%%%%%%%%%%%%%%%%%%%%%%%%%%

\subsection{Fake User and Bot Detection}

Not all messages come from human users of OSNs. There are programs known as \textit{bots} that generate and post messages regularly. Moreover, there are human-assisted or bot-assisted accounts called as \textit{cyborgs}. Separating bots and cyborgs from human users is important for user protection as well as for the online social network system to remove those bots. However, it should be noted that not all bots are malicious bots. There are legitimate bots for improving the usability of OSNs. Chu et al. \cite{chu-10} propose classification of human, bot, and cyborg accounts based on tweet content, tweeting behavior, and account properties. They propose a system composed of entropy-based, machine learning, account properties, and decision maker components.  The entropy-based component analyzes the inter-tweet interval. While a high value of entropy is an indicator of a bot (i.e., messages are sent at regular intervals), a low entropy value is a sign of a human user (i.e., irregular behavior). The machine learning component classifies tweet content as spam or not based on textual patterns. After crawling over 500,000 accounts, they have conducted their experiments on a balanced set composed of 1,000 bot, 1,000 cyborg, and 1,000 human accounts. The sensitivity values for human, cyborg, and bot classes are 0.949, 0.828, and 0.937, respectively. No human was misclassified as a bot, no bot was misclassified as a human in their experiments.

\textit{Fake users} may try to reach private information about OSN users. It is important to determine the intentions of users or at least to determine the likelihood of being a fake user. Social Privacy Protector (SPP) software \cite{fire-14} is a system developed to improve users' privacy and security on Facebook by identifying fake users. The SPP software has three layers of protection. In the first layer, the users who might pose a threat are categorized as fake users. The exposure of these possibly fake users to the user's profile is restricted. Each friend account of the user is processed and a credibility score is assigned to all friends based on a simple heuristics or a sophisticated algorithm benefiting from the degree of connection between the user and friends. The number of common friends and the number of pictures co-tagged are used to determine the strength of connections. Low scores indicate the likelihood of fake profiles. After identifying likely fake profiles, the basic privacy settings of Facebook are expanded in the second layer considering various types of social network profiles. The system keeps track of the user's internet activity and the applications that are installed. If any application has access to user's private information, a warning is issued to the user. The third layer analyzes, stores, and caches data for each user using the HTTP server. The main responsibility of the HTTP server is to connect SPP Firefox Add-on to the SPP Facebook application. SPP Add-on provides statistics about Facebook privacy settings and show vulnerability to fake profile attacks and third party applications. Each SPP user account and links to this account  are analyzed to generate the experiment data. The links of users are categorized into five groups of Facebook friends: 1) restricted friends by the user, 2) unrestricted friends by the user, 3) friends chosen to be restricted on purpose, 4) friends recommended for restrictions by the application  due to a low connection-strength score, and 5) friends not recommended for restrictions by the application. Based on this categorization, Facebook user's privacy settings are analyzed and classifiers are developed to determine fake Facebook profiles.

\subsection{Categorizing Social Network Actors}

Social networks are formed by groups of people that are linked by a social bond or relationship. There are different types of users such as individuals and organizations in social networks. While some of them are so active that they send message frequently and has many followers, some of them are generally inactive, but occasionally follow others. There are also users who send spam messages as mentioned in the previous section. Identifying types of users is important for security and privacy of users as well as marketing purposes.

In \cite{fazeen-11}, Twitter actors are categorized into four types: (a) leaders (who start tweeting and have many followers), (b) lurkers, who are generally inactive, but occasionally follow some tweets, (c) spammers, the unwanted tweeters, also called as twammers, and (d) close associates, including friends, family members, relatives, colleagues. Two classification approaches have been proposed to detect these social network (SN) actors: (a) context-dependent classification for situations where an abundant amount of tweet data is available and (b) context-independent classification (based on the actor tweet patterns) that is suitable when enough information is not available about the network context. Context-dependent classification method has two stages. In the first stage, the strengths of network links (or equivalently, actor-actor relationships) are evaluated with a fuzzy-set theoretic (FST) approach and a simple linear classifier that is used to separate the actor classes. The SN link (relationship) strengths  are estimated by using the followee-follower message statistics with fuzzy logic approach. Next, a large number of actors with strong social bonds (or link strengths) are eliminated. In the second stage, only tweeters with weak link strength (less than 15\% of maximum strength) are considered. A linear classification of  four actor types  is performed by using the number of tweets and the followee-follower ratio.

The second method performs actor classification by matching their short-term (roughly 25 days) tweet patterns based on the number of tweets with the generic tweet patterns of the prototype actors of different classes. The naive-Bayes (NB) classifier, multi-layer perceptron (MLP), and random forest (RF) classifier have been used for these tasks. Tweet patterns are constructed for the actor classes: the leaders (e.g., the news blogger), spammers, and associates. Since lurkers class has sparsely available data, it is not considered for this method. Given pattern of an actor is classified using patterns of actors of the classes. Experiments are conducted  on over 500 randomly sampled records from a Twitter database. According to their results, context-dependent classification method have a good separation of the four tweeter classes, leaders, lurkers, spammers, and associates. 

The passive users of social media, lurkers,  may follow the updates or content on Twitter but may not respond. Tagarelli et al. \cite{tagarelli2013s} propose centrality methods to rank the lurkers. The messages a user read in social networks (or posts as respond to him/her) can be considered as incoming links, whereas the messages he/she wrote (freely or as a response to others' messages) are considered as outgoing links. The naive approach to rank the users as a lurker is ranking them based on the ratio of incoming links/outgoing links. Besides the high number of incoming links/messages, the outgoing links should also be considered for defining lurkers. The problem in this approach is that it ignores who sends the messages and treats all the messages same. Messages coming from the other lurkers or messages from the influential people should not be treated equally. To do that, different metrics are proposed to rank the lurkers: in-neighbors-driven, out-neighbors-driven, in-out-neighbors-driven, etc. For each approach the formulas are created in such a way that it makes sure that the score of a node increases with the number of its in-neighbors and with their likelihood of being non-lurkers. And these metrics are evaluated by comparing with other approaches on  16M users and 132M links from Twitter data. Moreover, they report the top 20 lurkers for each method with their retweets and show that other methods fail to detect the lurkers and tag some of them as spams whereas proposed lurker ranking methods are able to detect them correctly. 

\subsection{Discussion}

Categorization of online social network users plays yet another crucial role in user characterization studies. This categorization can be performed using different techniques such as machine learning and honeypots in order to identify different user types such as spammers, fake users, bots, etc. These techniques mainly focus on the analysis of friend relations, content of online postings, posting behaviors, account properties, and so on.

In addition to the current state of the literature in spam and spammer (users sending frequent and unsolicited messages) detection, similarities and differences between users such as lurkers and spammers can be examined. Scores could be assigned to those users such that high scores could correspond to spammers whereas low scores could indicate lurkers. This could allow systems, public and private agencies, and businesses to comparatively classify the network users. Moreover, network access times and context interest could be analyzed together for detecting lurkers. Furthermore, boundary-spanning lurking can be also investigated to check if a person's lurking behavior change from one circle to another circle (i.e., lurker in one circle but active in another). For instance, the main idea of \cite{tagarelli2013s} is that people may have different access frequencies to their social networks and these access frequencies change with respect to time. Similarly, people tend to involve (e.g., send messages, comment and like) with the context they are interested in. These behaviors could be used for advertising and marketing purposes via recommendation systems as well as public/private investigations for various instances. Finally, the features from  \cite{stafford-13} (Table \ref{spammers_features}) and \cite{stringhini-10} (Table \ref{spam_message_features}) could be combined to study whether combined features increase the spam and spammer detection rate or not. Sakaura et al. \cite{sakakura-12} use the idea that spam messages will point to the same websites or resources and provide spammer detection for social bookmarking. This suggests that all links in messages could be crawled and analyzed to detect spam messages as spam links should occur frequently. 

Stafford et al. \cite{stafford-13} evaluate the effect of spams on twitter trending topics and conclude that spammers do not tend to drive trending topics on Twitter. It would be interesting to see the results of such research on other OSNs since they are also targeted by spammers. Rather than looking at the contents of postings, the links should be crawled to actually determine if it is spam or not. For example, ``Scientists identify remains as those of King Richard III http://t.co/kVGIV7to" may not look like a spam, but actually it could be a spam. 

As OSNs are also becoming target of fake users, bots and cyborgs for different purposes such as dissemination of viruses, criminal activities, personal investigations, cheating, etc. In addition to these categories, internet trolls also can be categorized by creating features which have been successfully applied to detect other user types. As discussed earlier, Chu et al. \cite{chu-10} and Fire et al. \cite{fire-14} utilize content of tweets, user behaviors and account properties. Those features can potentially be used for detecting ``social network trolls" as well. This could have broadened the target of the studies since trolls are real persons aim to provoke other users with hate speech, racism, controversial contents, etc. For example, Ortega et al. \cite{ortega-12} aim to identify propagation of trust and distrust for the detection of trolls in a social network. Detecting trolls with similar techniques used in \cite{chu-10,fire-14} could also be efficient and valuable along with the knowledge of other characters in OSNs.

%% file: 6matching_profiles.tex
\section{Matching Profiles for Online Social Networks} \label{sec:entitymatching}
There are many online forums providing different services and functionalities such as sharing ideas, meeting with friends, and following trends. While regular users of social networks are likely to have the same or similar profiles across online social networks, there are users who would like to represent themselves in a different way from one social network to another. The problem of matching profiles of the same person across different online social networks is usually studied as \textit{entity resolution}. For detecting these individuals, postings as well as profile and network information have been used in the literature. They extract features from profiles and determine the similarity of feature vectors to decide if they belong to the same person.

The entity resolution method proposed by Peled et al. \cite{peled2013entity} is based on machine learning methods. The probability that two user profiles from two different OSNs belong to the same individual is computed and ranked using features of the users extracted from their profiles. Their algorithm includes 4 steps composed of data acquisition, feature extraction, training set preparation, and  the classifier model construction. During data acquisition, user profiles are crawled and then irrelevant data are eliminated before further analysis. In the feature extraction step, they consider three different feature types which are name based features to compute name similarity, general user information based features such as location and employer, and social network topology such as the number of mutual friends.

String-based comparison measures and evaluation methods such as Soundex \cite{holmes-02}, longest common-substring \cite{friedman-92}, compression, difference, Jaro Winkler \cite{winkler-90}, Damareou Levenshtein distance \cite{navarro-01}, n-gram, VMN \cite{vosecky-09} are used for computing name similarities and name fequency similarity.
15 user information-based features are extracted to represent the similarities
between the personal information of two users. These features are 
 the distance of locations, current employer similarity using n-gram, Jaro Winkler and  Damareau Levenshtein metrics, Jaccard similarity, semi vector space model similarities, and vector space model of full profile similarity. After extracting features, users who have profiles on multiple OSNs are identified with comparing their names. Then, a manual check is performed by comparing the profile photos. Each pair is labeled as a match or no-match for setting up the training set. Users whose names do not match on multiple OSNs are treated as negative pairs. They have divided users into 10 groups randomly and removed labeled users that are not in the same group. Each group was tested using trained model using the other 9 groups. The authors evaluate their method on profiles from Facebook and Xing. They have collected 16,561 profiles from Facebook and 15,430 profiles from Xing. Among these profiles, 464 user profiles had the same first and last names (with 158 true matches and 306 false matches). They have utilized Rotation Forest \cite{rodriguez-06}, Random Forest \cite{Breiman-01}, Logistics Model Tree (LMT) \cite{Landwehr-05}, AdaBoost \cite{Freund-96}, LogitBoost \cite{Friedman-98}, and Artificial Neural Networks for building the classifier model. Although the name feature category provides the best results among all categories, the best score is obtained when all 27 features are combined to yield 0.98 AUC (area-under-the-curve) using LogitBoost.
 
Since OSNs such as Twitter limit the size of messages to be posted, this leads users to generate short messages. The style and choice of words can be used to identify the author of a message. Furthermore, the authorship or messaging patterns can be used for entity resolution. An Instant Messaging (IM) authorship analysis framework is proposed by Orebaugh et al. \cite{orebaugh-09,orebaugh2010data} to determine the entity of a cyber-criminal (i.e., the author of messages). Their main motivation is that people have patterns and they repeat these patterns. If the traces of these patterns can be determined and tracked, it could be possible to identify a criminal. Like people leave behavioral traces, they also leave textual traces that could be used to determine their identity. Such textual traces are useful for distinguishing people. So they  use IM conversation logs to identify and validate authors.  They propose classification methods to profile author behavior using various linguistics patterns and characteristics. They extract stylometric features of authors from text: syntactic and structural layout traits, patterns of vocabulary usage, unusual language usage, and stylistic feature. In the experiments, they gather data from IM conversation logs for four users and apply a series of preprocessing steps. J48 decision tree, IBk nearest neighbor, and Naive Bayes classifiers are used to determine if an author of an IM conversation could be identified based on his/her textual traces.	According to their results, Naive Bayes classification provide higher accuracy ($> 99\%$ accuracy for 4 users) than others. 

There are many reasons for using different aliases on OSNs. As some aliases may not be available when users join an OSNs, they are required to choose a new account name. Another reason for different aliases is not to reveal a person's true identity.
Johansson et al. \cite{johansson2013detecting} analyze string-based, stylo-metric-based, time profile-based, and social network based methods for detecting people with multiple aliases. String-based method matches aliases using Jaro-Winkler \cite{winkler-90} edit distance measure after normalizing the values to [0,1]. Time profile-based matching compares the number of postings made in a time window. More specifically, a normalized feature vector of the number of postings per hour is created for each user, and these feature vectors are compared by using Euclidean distance. Stylo-metric matching compares the statistical properties of messages such as the length of sentences and words, and the number of digits, punctuation characters (e.g.,
. ? ! , ; : ( ) " - '), 
and function words. Social network based matching considers threads or friends of user accounts. In this case, Jaccard similarity measure is used as the ratio of the number of common friends to the number of people who are friends to both aliases. For experimental analysis, the authors analyze 9 million documents collected from Irish Website forums. To be able to analyze all matching properties, they consider aliases having more than 60 posts. They have formed aliases into two equal-sized groups to perform alias matching, having 50 users up to 1000 users.
It has been observed that when the number of users increases, the accuracy decreases. 
Time profile-based matching outperforms stylometric based matching. The best performance is obtained when all matching techniques are applied together rather than applying each technique individually. Especially, combination of time profile-based and stylometric methods performs better than each individual matching technique. This combination provides 70\%, 55\%, and 43\% accuracies for 50, 250, and 1000 users, respectively.

Promising results on entity resolution have been obtained using messaging patterns, matching aliases, and profile and network similarity. It is difficult to compare the accuracies of these methods due to the number of users considered in evaluations.
Messaging patterns (syntactic and structural layout traits, patterns of vocabulary usage, unusual language usage, and stylistic feature)  can achieve $>99\%$ accuracy for 4 users using naive Bayesian classifier for authorship analysis \cite{orebaugh-09,orebaugh2010data}. %They  have observed that a person has a consistent style in IM conversations. 
Abbreviations and special characters are found to be the best discriminators. Especially, 'U', three dots and hyphen are detected as the strongest attributes. 
%Therefore, messaging behavior of users can be used as a feature to identify the author of the messages. 
Although matching aliases based on users' postings gives promising results for some users \cite{johansson2013detecting}, postings may not be reliable as the vocabulary of posting may vary based on a topic or an OSN. Hence, vocabulary or words appearing in postings of people who write on diverse topics may not be discriminating features for them. Therefore, additional features need to be extracted in addition to features from postings. Peled et al. \cite{peled2013entity} analyze name similarity, profile similarity and social network similarity for entity matching. While name feature only based on Jaro Winkler similarity yields 0.95 AUC, they obtain 0.98 AUC when all features are combined for a special training/testing scenario after dividing 464 users into 10 groups.  Johansson et al. \cite{johansson2013detecting} utilize similar types of features used by Peled et al. \cite{peled2013entity} (alias similarity, social network similarity and stylo-metric features) and add time-profile based feature that provides the number of postings in a time window. They have achieved  $70\%$ accuracy for 50 users.

\subsection{Discussion}
It would be good to analyze the set of features to determine the same users and to separate different users. Most likely, username and full name similarities may be good indicators of the same entity. Especially, these features would work well when users intend to use the same aliases on multiple OSNs but may have to make slight changes on their aliases in case the same alias is not available. However, username similarities may not work well if an individual may choose very dissimilar aliases (deliberately or not) or different people may use very similar usernames. Separate analysis of the accuracy of detecting different users with similar names and detecting the same users with different aliases (and names) would help to identify the critical features for entity matching.

Besides using alias information, full name, profile, social network topology, and posting time may also yield important information about users. Time profiling may provide information about favorite times of posting messages. However, it is possible that a person may use a professional OSN during working hours and an OSN of friends and relatives after working hours. In those cases, time profiling would not help much. Stylo-metric features may also change with respect to the OSN. In a professional community, the language could be formal but in an inner circle the language could be relaxed. Changing behavior of users in different contexts should also be analyzed by processing longer periods of data and large number of users.

%% file: 7conclusion.tex
\section{Open Research Issues and Conclusion}

The online social networks have enabled information sharing and increased online social communications at a tremendous level. We live in a millennium where we have new anxieties such as FOMO (fear of missing out). One of the biggest advantages of online social networks is the expression of speech and feelings without needing to reveal true identity. While online social networks had made a big contribution to freedom of speech, there also had been negative experiences of using online social networks. 

\subsection{Overview}
The explicit and implicit social connections over the online social networks may reveal many insights and analysis of the user characteristics and future actions of the entities.
In this paper, we targeted user characterization with the motivation of increasing benefits of using online social networks. The credibility of information is critical if actions will be taken based on them. The attributes such as gender, age, geo-location, profession can be provided false, and people can be victims of this false information. We have looked into research studies that analyze behaviors of online social media users such as motives for providing false information or withholding information. Users should know whom they interact with such as a real human, a bot, or a fake person. Identifying aliases of the same person on different networks may help to determine true identity of a person. Understanding mental models of users may help to develop better services for them and protect their privacy. The marketing may be targeted based on users' mental models by categorizing them based on the use of online social networks such as active users who post messages vs. users who just consume messages. We briefly summarize the current status of research below.

\textit{Identifying User Attributes.} N-grams or k-top features usually provide satisfactory results for gender detection. 
The highest accuracy that is reported for gender detection is $98.51\%$ \cite{Deitrick-2012} using modified Winnow neural network and feature selection from 2-grams and 1-grams.
In other research studies, the accuracies are $82.83\%$ for Enron email \cite{cheng-11}, $76.75\%$ for Reuters newsgroup \cite{cheng-11}, 
$82\%$ using first names \cite{alowibdi-13}, $85\%$ using combination of first name, user name, and color profiles \cite{alowibdi-13}. 
Gender detection has also been studied for various languages. 
Using k-top features $87\%$ and $83\%$ accuracy for gender detection is achieved 
for Turkish and Indonesian, respectively \cite{Ciot-13}. $76\%$ and $90\%$ accuracy is achieved for French using k-top features and sius grammatical construction, respectively \cite{Ciot-13}. 
For Japanese, $62\%$ accuracy using k-top features \cite{Ciot-13} and $85.75\%$ accuracy using bag-of-features \cite{ito-13} are achieved. Psycho-linguistic and gender-linked cues
increased accuracy around $10\%$ with respect to character based features when combined with other set of features \cite{cheng-11}.
For age detection, one issue is the number of age categories and their ranges for age group prediction. $40.9\%\pm 0.9$ accuracy is achieved for five age groups \cite{song-13}, and 
$88.2\%$ accuracy is achieved between min16 (ages from 11 to 15) and plus25 (25 and older) groups \cite{Peersman-2011}. For geo-location prediction, $49.9\%$ accuracy (within 100 miles) is achieved using only 250 words or less \cite{chang-12}.
More than $90\%$ accuracy is achieved for many professions on a social network \cite{wagner-13}. In general gender and age detection has been studied more than other user attributes. In the literature, the range of accuracies for gender detection is relatively wide (e.g., between around $62\%$ and $98.51\%$). If language dependent features
are not used, it would be good to evaluate the proposed techniques on others' datasets. For age detection, identifying age groups for prediction looks to be an important factor.
The research studies have been limited for other user attributes. Multi-attribute prediction may yield interesting results compared to prediction using attributes individually.

\textit{User Behavior Analysis.} For behavioral analysis, timing between messages, text length, and propagation are a good set of features for predicting personality traits \cite{adali2012predicting}.
The mean absolute error rate for predicting agreeableness and openness is around $11-12\%$. 
Neuroticism is predicted with an error rate of around $18-19\%$. The error rates for extroversion and conscientiousness error rates are between $14\%$ and $16\%$.
Messaging or tweeting behavior has been studied and significant accuracy improvement for users whose tweet rate (probability of tweeting one or more messages in a 10-minute window) is more than $0.2$ \cite{darmon2013predictability}
Classifying people's immediate reaction  when they learn their OSN accounts were hacked
has been achieved with $78.25\%$ accuracy \cite{zangerle-14}.
Analyzing the motives for deceptive behavior is still at its infancy stages \cite{griffin2012toward}. Having a social image or privacy concerns could be the reasons for providing false information.
Identifying users acting differently from their network could lead to users having malicious intentions \cite{savage-14}.
Determining sensitivity of attributes using privacy index measures may help users determine if their privacy is violated or not \cite{wang-13}.
In addition, a policy aware social miner tool that gives users of OSNs to restrict search results based on the categories of search or to refute search results is provided \cite{paradesi-12}.
Mazumder et al. develop a radicalization index using contents of tweets. They have generated a heat map of radicalization  using location index (or geo-tags) and validated their results
with the data of two institutes. Teoh et al. study how OSNs are treated at higher education and especially observe gender-based differences \cite{teoh2014gender}. These observations could help educators develop more effective ways of using OSNs at the educational institutions. 

\textit{Mental Models.} 
Learning the mental models of OSN users is important to provide more secure or more advanced communication interfaces \cite{blythe2012implementing}. Mental models are also important for behavior analyses of OSN users. In general, the common approach for scholars to analyze the mental models is to find the change through the group, time, place. Mental models influence user’s behavior inside or the outside of a group \cite{tamburrini2015twitter}. Twitter users change their vocabulary they use to their conversation partner when that partner is inside a group or outside of that group. Views, likes, and dislikes actions could be used to determine whether people have leadership qualities or not \cite{kohli2014modeling}. 
 A user may have different interests and opinions in different times \cite{grimaudo2014tucan}. This can be measured by text analyses with using text similarity techniques between different time intervals. Different geographical locations also means different mental models \cite{zheng-14}. Each OSNs user can be related to a (lat, lon) location and their message can be mapped to these locations. This approach not only shows the mental models of locations, but also enables to give a globalness/localness score for a given content.
Political preferences can be extracted by processing the user's messages in OSNs. 
Once a user's political preference found, political tendency of media outlets or other think-tank groups can be measured by its followers too \cite{Golbeck-11-a}.
Social network analysis metrics such as centrality, clique count and clustering coefficient, can be employed to investigate the gender based productivity and collaborations \cite{ozel2012link}. In a social science journal analysis, females found to be more collaborative than males and males' contribution and presence found more than females.

\textit{User Categorization.} As OSNs provide services for users to interact in a variety of ways and this interactions other users. In order to identify the OSN users/profiles as spammers and messages as spams, researchers create honey profiles using honeypots in OSNs. Total of 900 honey profiles deployed and activity logs of these profiles are recorded by \cite{stringhini-10}. Using six features these logs are analyzed and a total of 15,857 spammers are detected on Twitter. This number is exceeded in \cite{yang-14} by creating 1,512 honey profiles in shorter time. This yields 17,000 spammer accounts collected with 60\% accuracy. Wang et al. \cite{wang-14} propose a framework called SPADE in order to detect spam messages and spammers using 15 and 74 attributes, respectively. As a result, over 0.92 F-measure and $91\%$ detection accuracy on web page model (in cross domain classification), 0.87 F-measure on user profile model, and 0.89 F-measure on message models are achieved by SPADE. Naive Bayes classifier is used by Stafford et al. \cite{stafford-13} to predict spam messages by using 10 attributes and $90\%$ of the instances are classified correctly. Sakaura et al. \cite{sakakura-12} propose a scheme for the detection of social bookmarking attacks when spammers try modifying the search engine webpage ranking results. In addition to spammers, users are also categorized as fake users, bots, and cyborgs in \cite{chu-10} and \cite{fire-14} depending on the tweet contents, tweeting behaviors and account properties. In \cite{chu-10}, authors propose a system composed of entropy-based, machine learning, account properties, and decision maker components. Due to the high accuracy, no human was misclassified as a bot and no bot was misclassified as a human in the experiments. Social Privacy Protector (SPP) software \cite{fire-14} is a system developed to improve users' privacy and security on Facebook by identifying fake users. In addition to previous categorization studies, Fazeen et al. \cite{fazeen-11} classify actors in OSNs using two methods: 1) context-dependent approach and 2) context-independent approach. Finally, OSN users are categorized as lurkers and ranked by \cite{tagarelli2013s} using centrality methods based on incoming and outgoing messages of the users. This method consistently reached a ranking stability in the range of 35÷75 iterations in a short time. In addition to those, a recent study discusses the characteristics of contemporary, subtle social bots in \cite{ferrara14}. These social bots are interacting with humans and they have the capability of convincing people, manipulating ideas, and deceiving. All these studies commonly show that users may be interacting with real people with good or evil (malicious) characteristics as well as non-human users with deceptive behaviors. In order to keep our online world balanced, researchers must understand the behavior of these bots. We also believe that there are still other possible categories (trolls) that similar techniques can be applied to classify.

\textit{Matching Profiles for Online Social Networks.} Name based features are commonly used feature sets to match entities. Peled et al. \cite{peled2013entity} obtain 0.95 AUC using name based features. 
By adding profile information and social network topological based features they increase AUC to   0.98  using LogitBoost learning method. Messaging behavior of users yields $99\%$ accuracy using Naive Bayes classifier for 4 users \cite{orebaugh-09,orebaugh2010data}. When time profiling features are used in addition to stylometric features, the accuracy has increased from around $50\%$ to $70\%$ for 50 users \cite{johansson2013detecting}. Rather than using a single feature set, combination of  different types of features  provided the best results for entity matching. 

\subsection{Open Research Issues}
While significant progress has been made on user characterization for OSNs, the current research also reveals important future work to be conducted. After analyzing the existing research, we believe some of the areas for further work could be listed as follows:

\begin{itemize}
\item \textit{Identifying user attributes:}
\begin{itemize}
\item analyze bag of features along with n-grams and k-top features for user attribute detection,
\item incorporate full names (first, middle, and last names) and countries for gender detection,
\item use profile photos for gender detection in addition to other attributes,
\item analyze language preference with respect to both age and gender, and
\item perform research on other attributes (education level, socio-economic status, etc.),
\end{itemize}

\item \textit{User behavior analysis:}
\begin{itemize}
\item incorporate unbiased tweeting behavior (tweeting regardless of context) for predicting personality traits,
\item determine alternate ways for predicting neuroticism and extroversion when propagation is limited,
\item determine user groups for withholding information and deception,
\item build a model that would relate deceptive behavior and specific attributes,
\item build a model that would separate deception from unintentional mistakes,
\item support eye movement tracking for deception on OSNs supporting video chat,
\item identify motives for deceptive behavior (e.g., peer pressure, good social image, malicious intentions, etc.),
\item study the role of communities for deceptive behavior,
\item study deception with respect to behavioral models and personality traits,
\item develop frameworks for privacy management that merge technical and legal frameworks,
\item enhance and automate restriction and refutation framework for online searchers,
\item develop a push-based system for updates of refutations and restrictions,
\item study behavior of hacked users,
\item analyze radicalization behavior with respect to geo-location estimation in addition to geo-tags provided by OSNs, and
\item analyze OSN use with respect to gender considering personality traits.
\end{itemize}

\item \textit{Mental models:}
\begin{itemize}
\item{ develop new mental models for making decisions to follow/unfollow another user, approve friend request, the content (message) and the people to share,
}
\item{
analyze varying time intervals for proper analysis of interest change,
}
\item{
explore location (region) specific mental models as users of the same location may have common interests,
}
\item{ incorporate additional features for mental models (e.g., retweets rather than only follow information) such as like/dislike feedback in Twitter, and}
\item{perform discipline specific collaborations for gender based mental model analyses.}
\end{itemize}

\item \textit{User categorization:}
\begin{itemize}
\item examine and analyze similarities and differences between lurkers and spammers, 
\item study the access times or/and context interest for detecting a lurker for better results,
\item investigate boundary-spanning lurking (one person can be a lurker in a friend circle due to the variety of boundary limitations while being very active in another circle),
\item create features for ``social network trolls",
\item create link analyses and website crawlers for Twitter messages that would help to automate the search of spam messages and increase accuracy, and
\item identify the real people who create bot accounts,
\item study online social bots and find countermeasures before they become a danger for our online society,
\item develop new method/tools to detect and distinguish benign/malicious bots. 

\end{itemize}

\item \textit{Matching Profiles for Online Social Networks:}
\begin{itemize}
\item analyze accuracy of entity matching for the \textit{same} users
using \textit{different} usernames and names,
\item analyze accuracy of entity matching for \textit{different} users using the \textit{same} usernames and names,
\item analyze entity matching for multiple accounts of the same user in the same OSN, 
\item analyze user behavior for entity matching with respect to multiple OSNs, and 
\item increase the number of users in the datasets and perform analysis for longer periods of data collection to consider changing behavior.
\end{itemize}

\end{itemize}

A large and diverse number of features have been studied for attribute detection, behavior analysis, mental models, user categorization, and entity matching. It would be good to have a feature extraction tool that could derive all the mentioned features (e.g., profile, name, color, linguistic, stylo-metric, etc.) in the literature. Once all these features are available, feature reduction tools such as principle components analysis or feature selection could be applied to determine the best set of features. Moreover, deep learning can benefit for increasing the accuracy of predictions. It would be good to study how deep learning increases prediction of gender, fake users, and personality traits. Accuracy measure could be easily misinterpreted if the original data and the number of categories are ignored. A plain comparison of accuracy values is misleading. Other accuracy measures should be utilized and confusion matrix should be provided wherever possible.
As another future direction,  quick feedback to OSNs might be critical in some scenarios, and Big Data technologies could be utilized for real-time alerts.

\subsection{Concluding Remarks}
The comprehension of characteristics of the user behavioral patterns in online social networks is significant  in the sense that these patterns can give useful information not only about the online interactivity but also about the people behaviors. These observations are crucial in order to  address the major personal problems from the perspective of psychological and sociological health of the individuals as well as making the society a better environment by understanding the individuals who live in it.  

Social network analysis is a very broad research area. The available social network APIs enable to crawl data and perform research on various aspects of social networks. In this paper, our goal was to categorize user characterization papers. We especially considered user attributes, behavior, mental models, categorization, and entity resolution. There are also other relevant research that can benefit user characterization. The papers covered in this survey help us to categorize user characterization research and help other researchers provide a guide of exploring new ideas about user characterization.